\documentclass[trackchanges,twocolumn]{aastex701}
\usepackage{makecell}
\usepackage{graphicx}

\begin{document}

\title{AT\,2022csn: A Photometrically Peculiar Optical/UV Tidal Disruption Event in a Type II AGN}

\author[orcid=0000-0002-7579-1105]{Yael Dgany}
\affiliation{School of Physics and Astronomy, Tel Aviv University, Tel Aviv 69978, Israel}
\email[show]{yaeldgany@gmail.com}  

\author[orcid=0000-0001-7090-4898]{Iair Arcavi} 
\affiliation{School of Physics and Astronomy, Tel Aviv University, Tel Aviv 69978, Israel}
\email{arcavi@gmail.com}

\author[0000-0002-7817-0099]{Grisha Zeltyn}
\affiliation{School of Physics and Astronomy, Tel Aviv University, Tel Aviv 69978, Israel}
\email{grishazeltyn@tauex.tau.ac.il}

\author[0000-0001-6395-6702]{Sebastian Gomez}
\affiliation{University of Texas at Austin, 1 University Station C1400, Austin, TX 78712-0259, USA}
\email{sebastian.gomez@austin.utexas.edu} 

\author[0000-0002-1125-9187]{Daichi Hiramatsu}
\affiliation{Department of Astronomy, University of Florida, 211 Bryant Space Science Center, Gainesville, FL 32611-2055 USA}
\email{dhiramatsu@ufl.edu} 

\author[0000-0002-3683-7297]{Benny Trakhtenbrot}
\affiliation{School of Physics and Astronomy, Tel Aviv University, Tel Aviv 69978, Israel}
\email{benny.trakht@gmail.com}

\author[0000-0001-9570-0584]{Megan Newsome}
\affiliation{University of Texas at Austin, 1 University Station C1400, Austin, TX 78712-0259, USA}
\email{newsome.megane@gmail.com}

\author[0000-0003-4974-3481]{Dalya Baron}
\affiliation{Kavli Institute for Particle Astrophysics \& Cosmology, Stanford University, CA 94305, USA}
\affiliation{Center for Decoding the Universe, Stanford University, CA 94305, USA}
\email{dalyabaron@gmail.com}

\author[0000-0002-7472-1279]{Craig Pellegrino}
\affiliation{NASA Goddard Space Flight Center, 8800 Greenbelt Road, Greenbelt, MD 20771, USA}
\email{craig.m.pellegrino@nasa.gov}

\author[orcid=0000-0003-0035-6659]{Jamison Burke}
\affiliation{Shady Side Academy, 423 Fox Chapel Road, Pittsburgh, PA 15238, USA}
\email{jamburke@shadysideacademy.org,}

\author[0000-0003-0901-1606]{Nadejda Blagorodnova}
\affiliation{Institut de Ci\`encies del Cosmos (ICCUB), Universitat de Barcelona (UB), c. Mart\'i i Franqu\`es, 1, 08028 Barcelona, Spain} 
\affiliation{Departament de F\'isica Qu\`antica i Astrof\'isica (FQA), Universitat de Barcelona (UB), c. Mart\'i i Franqu\`es, 1, 08028 Barcelona, Spain} 
\affiliation{Institut d'Estudis Espacials de Catalunya (IEEC), Edifici RDIT, Campus UPC, 08860 Castelldefels (Barcelona), Spain}
\email{nblago@fqa.ub.edu}

\author[0000-0003-3947-5946]{David Jones}
\affiliation{Instituto de Astrof\'isica de Canarias, E-38205 La Laguna, Spain} 
\affiliation{Departamento de Astrof\'isica, Universidad de La Laguna, E-38206 La Laguna, Spain} 
\email{david.jones@iac.es}

\author[0009-0001-4098-7706]{Gerard Garcia-Moreno}
\affiliation{Institut de Ci\`encies del Cosmos (ICCUB), Universitat de Barcelona (UB), c. Mart\'i i Franqu\`es, 1, 08028 Barcelona, Spain} 
\affiliation{Departament de F\'isica Qu\`antica i Astrof\'isica (FQA), Universitat de Barcelona (UB), c. Mart\'i i Franqu\`es, 1, 08028 Barcelona, Spain} 
\email{ggarcimo@fqa.ub.edu}

\author[0000-0001-8985-2493]{Erez A. Zimmerman}
\affiliation{Department of Particle Physics and Astrophysics, Weizmann Institute of Science, 234 Herzl Street, 7610001 Rehovot, Israel}
\email{erezimm@gmail.com}

\author[0009-0000-6748-4319]{Asaf Horowicz}
\affiliation{Department of Particle Physics and Astrophysics, Weizmann Institute of Science, 234 Herzl Street, 7610001 Rehovot, Israel}
\email{asaf.horowicz@weizmann.ac.il}

\author[0000-0002-3653-5598]{Avishay Gal-Yam}
\affiliation{Department of Particle Physics and Astrophysics, Weizmann Institute of Science, 234 Herzl Street, 7610001 Rehovot, Israel}
\email{avishay.gal-yam@weizmann.ac.il}

\author[0000-0002-1650-1518]{Mariusz Gromadzki}
\affiliation{Astronomical Observatory, University of Warsaw, Al. Ujazdowskie 4, 00-478 Warszawa, Poland}
\email{mariusz.gromadzki@gmail.com}

\author[0000-0002-9392-9681]{Edo Berger}
\affiliation{Center for Astrophysics \textbar\ Harvard \& Smithsonian, 60 Garden Street, Cambridge, MA 02138--1516, USA}
\affiliation{The NSF AI Institute for Artificial Intelligence and Fundamental Interactions, USA}
\email{eberger@cfa.harvard.edu}

\author[0009-0007-8485-1281]{Sara Faris}
\affiliation{School of Physics and Astronomy, Tel Aviv University, Tel Aviv 69978, Israel}
\email{sarafaris452@gmail.com}

\author[0000-0003-4253-656X]{D. Andrew Howell}
\affiliation{Department of Physics, University of California, Santa Barbara, CA 93106-9530, USA}
\affiliation{Las Cumbres Observatory, 6740 Cortona Drive, Suite 102, Goleta, CA 93117-5575, USA}
\email{ahowell@lco.global} 

\author[0000-0003-0871-4641]{Harsh Kumar}
\affiliation{Center for Astrophysics \textbar\ Harvard \& Smithsonian, 60 Garden Street, Cambridge, MA 02138--1516, USA}
\affiliation{The NSF AI Institute for Artificial Intelligence and Fundamental Interactions, USA}
\email{harsh.kumar@cfa.harvard.edu}

\begin{abstract}
The emission mechanism and host galaxy preference of optical/UV tidal disruption events (TDEs) are still not entirely understood. We present observations of the TDE AT\,2022csn, which is one of the most distant ($d_L\sim$726\,Mpc) and luminous ($L_{ peak}=2.487^{+0.073}_{-0.067}\times10^{44}\,{\rm erg\, s^{-1}}$) optical/UV TDEs observed to date. Although it is a spectroscopically normal H+He TDE, it shows some photometric peculiarities, exhibiting a pronounced double-peaked light curve (with peaks separated by 18.30 $\pm$ 2.84 days in the $g$-band), and lying in the low-temperature and large-radius end of the optical/UV TDE population. 
The host galaxy of AT\,2022csn shows evidence for a significant starburst within the last $\sim$Gyr consistent with other optical/UV TDEs, but also narrow emission lines that place it within the Type II AGN region of the BPT diagram. Interaction between the TDE and a pre-existing AGN accretion disk might explain the peculiar photometric properties. However, it is puzzling that a TDE would be visible in a Type II AGN, where according to the AGN unification picture the central region around the supermassive black hole is obscured. We suggest a few scenarios to reconcile this. AT\,2022csn together with AT\,2019ahk, which shows similar properties, may belong to a new subset of low-temperature, high-radius TDEs in galaxies with Type II AGN emission features.

\end{abstract}


\section{Introduction}
Supermassive black holes (SMBHs) likely reside at the centers of all massive galaxies in the local Universe. While a fraction of these SMBHs are observed as active galactic nuclei (AGNs), in which ongoing accretion produces luminous emission, the majority are quiescent and therefore difficult to study directly \citep[e.g][]{Green_and_Ho_2007,Mullaney_2013}. One of the few observational probes of distant inactive SMBHs is the transient emission produced during a tidal disruption event (TDE). A TDE occurs when a star passes sufficiently close to an SMBH that tidal forces overcome the star’s self-gravity leading to its disruption \citep{Hills_1975}. Specficially, when the pericenter of the star's orbit is equal or lower than the tidal radius, $R_t\approx(M_{BH}/M_\ast)^{1/3}R_\ast$ (where $M_\ast$ and $R_\ast$ are the mass and radius of the star respectively, and $M_{BH}$ is the mass of the SMBH).

In the case of a full disruption (in which the star is completely torn apart), roughly half of the debris remains gravitationally bound to the SMBH and returns to it at a rate $\dot{m}\propto t^{-5/3}$, while the other half becomes unbound and escapes the system \citep{Rees1988Natur.333..523R,phinney1989IAUS..136..543P,Kochanek_1989}.
For solar-mass stars disrupted by non-rotating SMBHs with masses $\lesssim10^8\,M_{\odot}$, the tidal disruption will occur outside the event horizon, allowing for an observable accretion flare. 

TDE flares can thus confirm the presence of an SMBH and provide a promising mechanism to constrain its mass and potentially its spin \citep[e.g.][]{Leloudas_2016NatAs...1E...2L}. As such, TDEs offer a window into the SMBH population and can, in addition, help constrain accretion physics. However, these events are relatively rare with rates in
the range of $10^{-5}$--$10^{-4}$  events per galaxy per year \citep[e.g.][]{Wang_2004,Stone_and_Metzger_2015}.

Theoretical predictions of TDE emission anticipated that TDEs would radiate primarily in X-rays from directly observed compact accretion disks. Indeed, the first TDEs were discovered with the ROSAT satellite \citep{ROSAT1999A&A...349..389V} as luminous, soft X-ray flares in the nuclei of otherwise quiescent galaxies (e.g. \citealt{Bade_1996,Komossa_1999,Grupe_1999,Greiner_2000}; see \citealt{Saxton_2021} for a review). 

Surprisingly, in recent years a distinct population of TDEs discovered primarily at optical and ultraviolet (UV) wavelengths has emerged (e.g. \citealt{Gezari_2006,Van_velzen_2011ApJ...741...73V,Gezari_2012,Arcavi_2014}; see \citealt{van_Velzen_2020} for a review). 
Optical/UV TDEs typically reach peak optical absolute magnitudes of $\sim$ -20 and rise on timescales of weeks \citep[e.g.][]{van_Velzen_2020,van_Velzen_2021}. 
These events are mostly blue, with blackbody temperatures
of a few $10^4$ K and blackbody radii of several $10^{-4}$ pc lasting for several months to years
\citep[e.g.][]{Van_velzen_2011ApJ...741...73V,Gezari_2012,Arcavi_2014,van_Velzen_2020}. In some cases, the bolometric luminosity follows a decline consistent with the canonical $t^{-5/3}$ mass fallback rate. 
Some optical/UV TDEs are accompanied by X-ray emission with properties that vary strongly between events \citep[e.g.][]{Holoien_2016,Saxton_2021,Liu_2022,Guolo_2024}. 

A few mechanisms have been proposed to explain the origin of the optical/UV emission in TDEs. In the first, X-ray emission from a rapidly circularized accretion disk is reprocessed by optically thick material  surrounding the disk \citep[e.g.][]{Guillochon_2013,Roth_2016,Dai_2018ApJ...859L..20D}. In the second, the optical/UV emission is attributed to shocks as the returning stellar debris circularize through collisions at apocenter due to relativistic precession \citep{Piran_2015}. 
Some studies \citep[e.g.][]{Lu_Bonnerot_10.1093/mnras/stz3405,Steinberg2024} suggest an intermediate scenario in which circularization begins near pericenter, while the emission powering the light-curve peak is dominated by stream-disk shocks that further circularize the debris. 
An additional scenario attributes the optical/UV emission to radiative cooling of a hot, extended envelope formed during prompt circularization \citep{Metzger_2022}.

Spectroscopically, in addition to their characteristically strong blue continuum, TDEs commonly exhibit broad (FWHM $\sim$ 10,000\,km\,s$^{-1}$) \ion{He}{2} $\lambda$4686 emission \citep{Gezari_2012,Arcavi_2014}, which is not observed with comparable widths, strengths, or persistence in other known classes of transients.
In some cases broad Balmer lines are present as well \citep[e.g.][]{Arcavi_2014,Gezari_2015ApJ...815L...5G,Hung_2017ApJ...842...29H}. These events are commonly classified as H-, He-, or H+He- TDEs, according to their dominant emission line species. However, the physical origin of these spectral properties remains an active topic of debate.
\citet{charalampopoulos2022detailed} found that the the ratio between \ion{He}{2} and H$\alpha$ emission increases as the inferred blackbody radius decreases, likely as a result of optical depth effects. This behavior suggests a stratified photosphere, consistent with some theoretical expectations \citep{Roth_2016}.
The large observed line widths have been interpreted as either Doppler broadening \citep{Ulmer_1999ApJ...514..180U,Guillochon_2013} or as dominated by electron scattering wings \citep{Roth_2018ApJ...855...54R}. 

Apart from the uncertainties surrounding their emission mechanism, optical/UV TDEs show a strong and not fully understood preference for post-starburst host galaxies \citep{Arcavi_2014,French_2016,French_2020}. This preference may be linked to the dynamical structure of galaxy nuclei several 100 million years after strong star formation episodes \citep{French_2020_host_structure}. Recently,  \cite{Newsome_2025} found that the post-starburst history in some TDE hosts may be radially stratified, with younger stars concentrated closer to the nucleus. This is consistent with inward-propagating starburst, possibly driven by gas inflow that may contribute to the enhanced TDE rate. Studies of optical/UV TDEs therefore also provide a window into nuclear stellar dynamics and various processes that can enhance TDE rates in post-starburst environments \citep[see also][for another possible TDE rate enhancement mechanism]{Madigan2018ApJ...853..141M}.

To date, TDE searches and samples have largely focused on quiescent or weakly active host galaxies, with AGN hosts often excluded due to the observational challenges of separating TDEs from AGN activity \citep[e.g.][]{Hung_2018,van_Velzen_2020,Needle_in_haystack2023ApJ...957...57D}. However, TDEs should still occur in AGN. In fact, the TDE rate may even be enhanced in such environments \citep{Stone_2018,Kaur_2025ApJ...979..172K}, a possibility that may also be supported by recent discoveries of extended emission-line regions in post-starburst galaxies that hosted TDEs \citep{Wevers_2024,Pursiainen_2026}. TDEs in AGN may also exhibit distinct observational signatures \citep{Chan_2019}, such as in the case of 1ES 1927+654 (\citealt{Trakhtenbrot_2019,Ricci_2020}; but see \citealt{Scepi_2021} for an alternative interpretation). TDEs in AGN offer a valuable opportunity to study the interplay between stellar disruption, pre-existing accretion flows, and the circumnuclear medium. 

AT\,2019ahk \citep{Holoien_2019} was a H  TDE in a host galaxy that likely harbors a Type II AGN. Type II AGNs are characterized by strong narrow-line emission with no broad-line components. In the AGN unification picture \citep{unified_model1993ARA&A..31..473A}, the broad-line region (BLR) is closer to the SMBH than the narrow-line region (NLR), and so in AGN with no broad lines the central region around the SMBH is thought to be obscured by a dusty torus. As the broad lines seen in TDEs are roughly of the same width (or wider) than those seen in broad-line AGN (and hence could be emitted from the same region), this raises the question of how a TDE could be observed in systems where the BLR is obscured (as the same torus should also obscure the TDE).

In this work, we present AT\,2022csn, a spectroscopically-normal H+He TDE showing peculiar photometric properties, and which occurred in a likely Type II AGN.
We describe our observations in Section \ref{section:observations}, their analysis in Section \ref{section:analysis}, the host galaxy analysis in Section \ref{section:host_galaxy}, discuss our results in Section \ref{section:discussion}, and
summarize in Section \ref{section:summary}.
We adopt a flat $\Lambda$CDM cosmology with parameters from \cite{Planck182020A&A...641A...6P}, namely $H_0=67.66\ {\rm km\ Mpc^{-1}\ s^{-1}},\ \Omega_{m,0}=0.30966$ and $\Omega_\Lambda = 0.69034$.

\section{Discovery and Classification} 

AT\,2022csn was discovered by the Zwicky Transient Facility (ZTF; \citealt{ZTF2019PASP..131a8002B}) as ZTF22aabimec on 2022 February 19, 07:52:19 (UTC used throughout), which is MJD 59629.33, with a \textit{g}-band magnitude of $19.00\pm0.21$. At RA = 112.228874 and Dec = +26.897031 \citep{fremling2022TNSTR.452....1F}, it is consistent with the center of the galaxy SDSS J072854.92+265349.3, as retrieved via the NASA/IPAC Extragalactic Database (NED). \cite{Needle_in_haystack2023ApJ...957...57D} identified the event as a possible TDE in a systematic search for transients in galaxy centers using the Lasair broker \citep{lasair10.1093/rasti/rzae024}.
AT\,2022csn was also detected by the Gaia photometric science alert team \citep{Gaia2021A&A...652A..76H} as Gaia22ayp, the Panoramic Survey Telescope and Rapid Response System (PanSTARRS; \citealt{PanSTARRS2004SPIE.5489...11K}) as PS22bju and by the Asteroid
Terrestrial-impact Last Alert System (ATLAS; \citealt{ATLAS_FP2_2020PASP..132h5002S}) as ATLAS22ggz.

A spectrum of AT\,2022csn was obtained on 2022 March 09 by the extended Public ESO Spectroscopic Survey for Transient Objects (ePESSTO+) and was reported to the Transient Name Server \citep[TNS;][]{TNS2021AAS...23742305G}. It showed a blue continuum with narrow [\ion{O}{3}] $\lambda5007$ and H$\alpha$ emission lines, prompting the classification of AT\,2022csn as a Type I superluminous supernova (SLSN-I) at redshift $z = 0.15$ \citep{2022TNSCR.667....1S}. 

Subsequently, on 2022 December 25, \citet{2022TNSCR3660....1A} reclassified it as a H+He TDE as it developed broad \ion{He}{2} features. The redshift was refined to $z = 0.148$ (derived from narrow H$\alpha$ and [\ion{O}{3}] $\lambda\lambda 5007,4959$ emission features), implying a luminosity distance of 726.20 Mpc.

\section{Observations}\label{section:observations}

\subsection{Photometric Observations}\label{Section:photometric observations}

We obtained \textit{UBVgri}-band photometry beginning on MJD 59639 through the global 1m telescope network of the Las Cumbres Observatory \citep{LasCumbres2013PASP..125.1031B}. Reference images were acquired from Las Cumbres on MJD 60270, over one year and a half after discovery.
To remove host-galaxy emission, we used the \texttt{lcogtsnpipe} pipeline \citep{lcogtsnpipe10.1093/mnras/stw870}, which generates the Point Spread Function (PSF) for each image and employs an implementation of the High Order Transform of PSF ANd Template Subtraction (\texttt{HOTPANTS}; \citealt{hotpants2015ascl.soft04004B}) algorithm to perform image subtraction, followed by PSF photometry at the source position.
\textit{UBV}-band photometry was calibrated using standard Landolt fields observed on the same nights as the TDE field, while \textit{gri}-band photometry was calibrated to the AB system using the Sloan Digital Sky Survey (SDSS) Data Release (DR) 14  \citep{DR14Abolfathi_2018}. 

We retrieved additional PSF-fit photometry of reference-subtracted images from the ZTF forced photometry service \citep{ZTF_FP2019PASP..131a8003M} between MJD 58156 and 60774, and from the ATLAS forced photometry service 
\citep{ATLAS_FP2018PASP..130f4505T,ATLAS_FP2_2020PASP..132h5002S} between MJD 57314 and 60772 at the position of AT\,2022csn. 

Four epochs of target-of-opportunity observations were obtained (PI: Schulze) with the Ultra-Violet/Optical Telescope (UVOT; \citealt{UVOT2005SSRv..120...95R}) and the X-Ray Telescope \citep[XRT;][]{Burrows2005} on board the Neil Gehrels Swift Observatory \citep[hereafter, Swift;][]{Gehrels_2004} between MJD 59658 and 59671. We (PI: Dgany) obtained an additional epoch for host-galaxy photometry on MJD 60758. 
We extracted photometry from the UVOT images using the uvotsource task of the NASA High Energy Astrophysics Science Archive Research Center (HEASARC) \texttt{HEASoft} software package. We performed aperture photometry using a 7\arcsec\ aperture for the source and a 37.5\arcsec\ aperture for the background in the UV bands, while a 5\arcsec\ aperture was used for the source in the $b$ and $v$ bands due to the proximity of another optical source. On the fourth epoch, the position of the TDE fell on an area of low detector sensitivity and was masked out by the small scale sensitivity check performed by \texttt{HEASoft} in the $uvw1$, $uvw2$, and $uvm2$ bands. We thus discard these images. 

\begin{figure*}[ht]
\includegraphics[width=\textwidth]{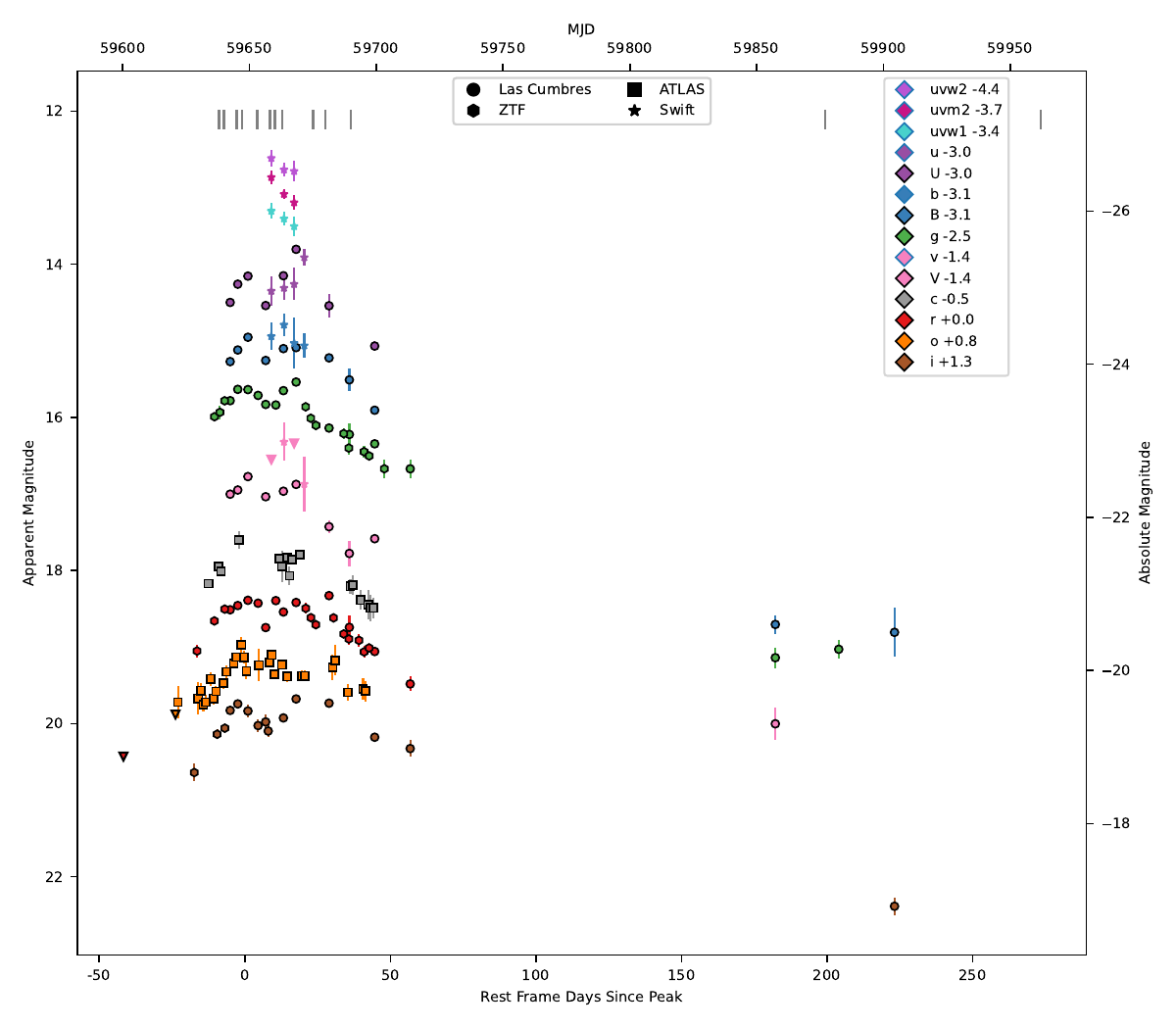}
\caption{Host subtracted Milky Way extinction corrected light curve of AT\,2022csn. Vertical grey lines mark the times of spectroscopic observations during the TDE (i.e. not including the host-galaxy spectra). Triangles indicate non-detection upper limits, corresponding to $3\sigma$ for the Swift measurements and $5\sigma$ for the other sources.}
\label{fig:full_lightcurve}
\end{figure*}

We correct all optical and UV photometry for Milky Way extinction assuming a \citet{Cardelli_1989}
extinction law with $R_V$ = 3.1 and a line of sight reddening of  of $E(B-V) = 0.058$, as retrieved from NED using the \citet{2011ApJ...737..103S} extinction map. The optical and UV photometry is provided in Table \ref{tbl:photometric_measurements} and shown in Figure \ref{fig:full_lightcurve}. 

Swift XRT data were processed using \texttt{XIMAGE}. No significant X-ray emission was found at the position of AT\,2022csn. We used \texttt{XIMAGE} to calculate flux upper limits for each XRT epoch, using a circular aperture with a radius of  47\arcsec\, centered on the optical position of the transient, after checking that no other X-ray sources were within the aperture. Assuming a Galactic column density of $n_{\rm H} = 6.27 \times 10^{20}\,{\rm cm^{-2}}$ and a power-law spectrum with photon index $\Gamma=2$ \citep{Ricci_2017}, we derive $3\sigma$ upper limits for the absorbed flux ranging from $1.203$ to $2.650\times10^{-13}\,{\rm erg\,s^{-1}\,cm^{-2}}$, which correspond to an average luminosity of $L_{0.2-10\,{\rm keV}} < 1.21\times10^{43}\,{\rm erg\,s^{-1}}$. When combining all XRT epochs (with a total exposure time of approximately 8.4 ks), we still find no significant X-ray detection to a combined $3\sigma$ upper limit of $L_{0.2-10\,{\rm keV}} < 3.8 \times 10^{42}$ ${\rm erg\,s^{-1}}$.

\begin{center}
\begin{figure}[ht]
\includegraphics[width=0.4\textwidth]{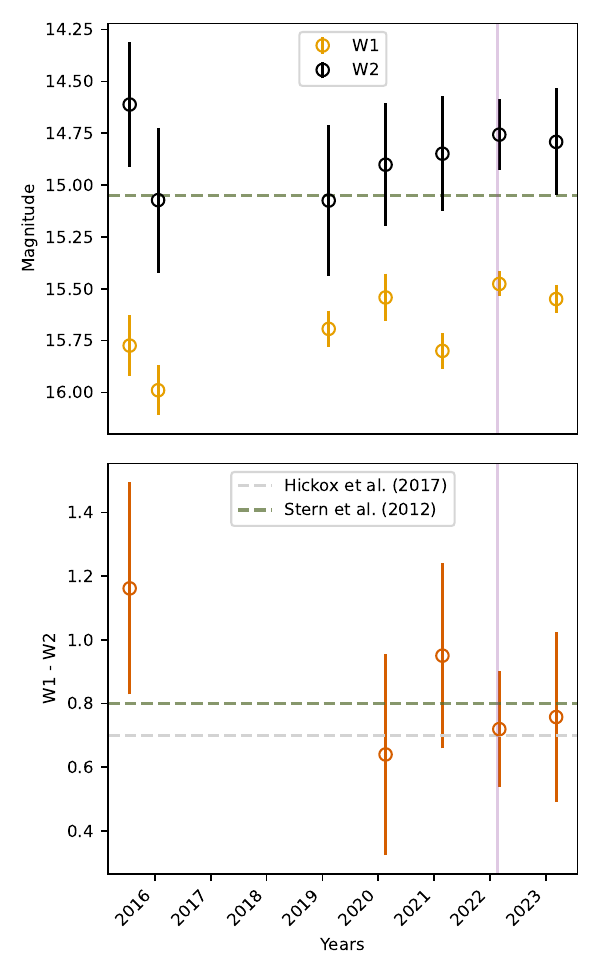}
\caption{NEOWISE $W1$ and $W2$ photometry (top) and color (bottom) at the position AT\,2022csn. The dashed horizontal line in the top panel marks the $W2=15.05$ mag threshold adopted from \citet{NEOWISE_0_8_2012ApJ...753...30S}, below which we do not calculate $W1-W2$ colors. AGN MIR color thresholds from \citet{Hickox_2017} and \citet{NEOWISE_0_8_2012ApJ...753...30S} are indicated as dashed horizontal lines in the bottom panel. The discovery date of AT\,2022csn is indicated by a vertical purple line.}
\label{fig:NEOWISE}
\end{figure}
\end{center}

We searched the Near-Earth Object Wide-field Infrared Survey Explorer (NEOWISE; \citealt{Mainzer_2011_NEOWISE}) database via the NASA/IPAC Infrared Science Archive \citep{NASA_IPAC2018cwla.conf...25T} for mid-infrared (MIR) detections located within 1\arcsec\, of the position of AT\,2022csn. NEOWISE acquires multiple exposures of each target during semiannual observing phases. Through this search, we obtained data covering the period from MJD 57116 to 60385. The observations were processed using a custom Python script that kept sources with photometric quality flags $ph_{qual}$ = A--C, no contamination ($cc_{flags}$ = 0) and required the detections to be unblended (nb = 1, na = 0). Points with poor PSF fits (rchi2 $\geq$ 2.5) and SNR below 3 were removed. We correct the WISE MIR photometry using the \citet{Fitzparik_1999PASP..111...63F} extinction law with the corresponding coefficients from
\citet{Yuan_2013}.
We estimate the host-galaxy MIR flux from all pre-TDE epochs as the mean flux and its variance, subtracted this baseline from all post TDE discovery epochs in flux space, and found no residual emission exceeding the $3\sigma$ level in either band.
The unsubtracted weighted-mean magnitudes for each NEOWISE visit and corresponding  
$W1$ - $W2$ colors are shown in Figure \ref{fig:NEOWISE}.

We also retrieved PanSTARRS DR1 \textit{grizy} magnitudes \citep{PS1_2017yCat.2349....0C} of the host galaxy obtained on MJD 56025 from VizieR\footnote{\url{https://vizier.cds.unistra.fr/viz-bin/VizieR}} \citep{vizier,vizier2000} as well as GALEX \textit{FUV} \citep{GALEX2005ApJ...619L...1M} and DES \textit{rz} host-galaxy photometry \citep{DES2016} from the \texttt{Blast}\footnote{\url{https://blast.scimma.org/}} database \citep{jones2024blastwebapplicationcharacterizing} and SDSS \textit{ugriz} \citep{Abazajian_2009} obtained from NED.
In addition, we used the Infrared Astronomical Satellite \citep[IRAS;][]{IRAS_1984} Scan Processing and Integration tool (Scanpi) to stack far-infrared IRAS scans at the position of the host galaxy, from which we derived $5\sigma$ upper limits at 60$\mu$m and 100$\mu$m.

\begin{deluxetable*}{llllll}
\tablecaption{AT\,2022csn optical and UV photometry (not corrected for Milky Way extinction).}
\tablehead{\colhead{MJD}  & \colhead{Mag} & \colhead{Err} & \colhead{Filter} & \colhead{Source} & \colhead{System}}
\startdata
59620.79 & $>$19.21 & ... & $o$ & ATLAS & AB \\
59645.85 & 18.310 & 0.010 & $g$ & Las Cumbres & AB\\
59656.86 & 18.481 & 0.015 & $B$ & Las Cumbres & Vega \\
59658.11 & 16.96 & 0.10 & $uvw1$ & Swift & Vega \\
59687.16 & 18.96 & 0.055 & $r$ & ZTF & AB \\
\enddata
\tablecomments{The complete table is available in machine-readable format. A subset is presented here to illustrate its structure and content.}
\end{deluxetable*}\label{tbl:photometric_measurements}

\subsection{Spectral Observations}
\begin{figure}[ht]
    \includegraphics[width=0.5\textwidth]
    {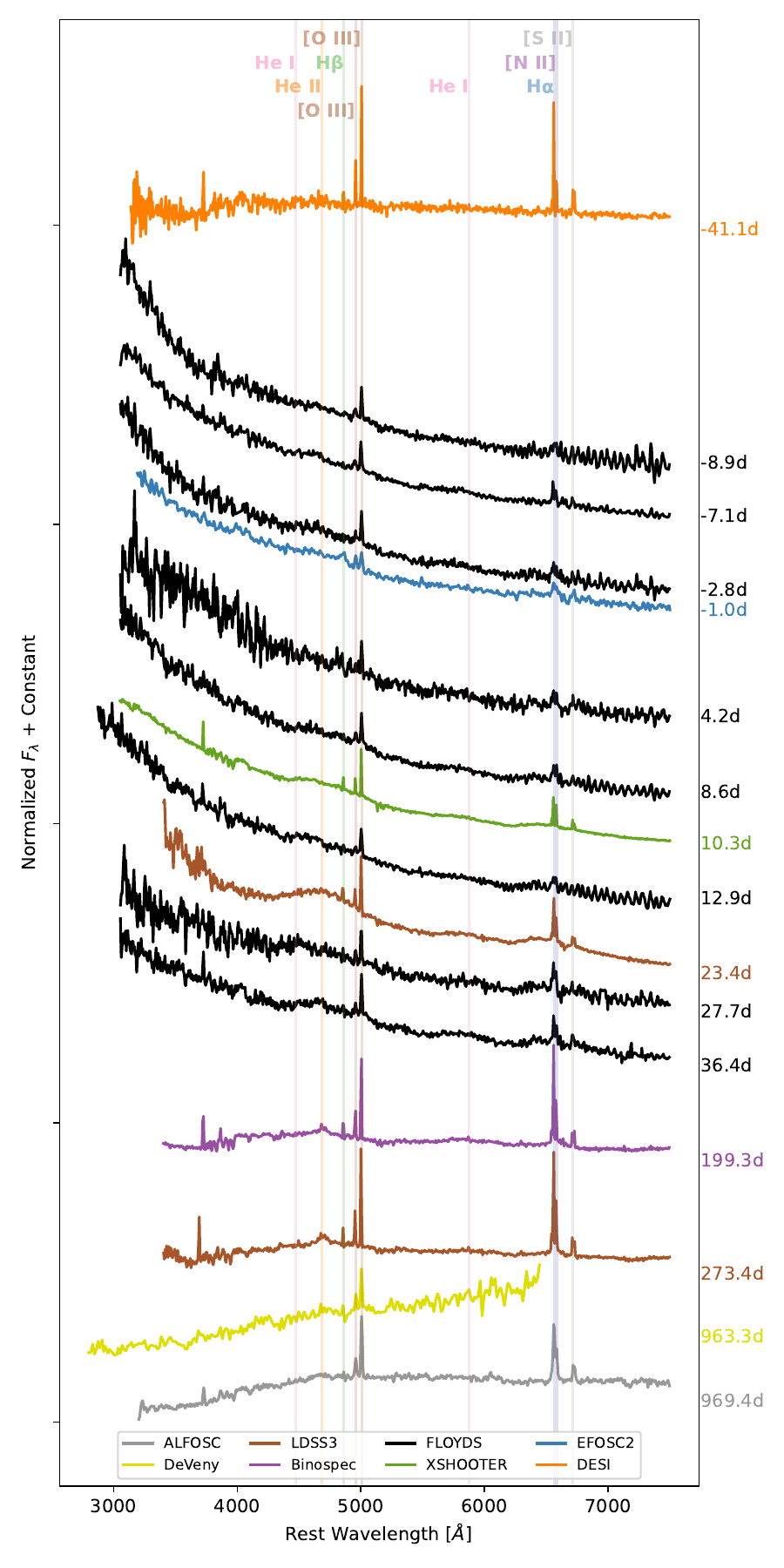}
    \caption{Photometry-calibrated spectra of AT\,2022csn and spectra of its host galaxy from both before and after the flare. Rest frame days from the first $g$-band peak are marked.}    \label{fig:Stacked_spectra_norm}
\end{figure}
We obtained optical spectroscopy of AT\,2022csn using the FLOYDS spectograph \citep{FLOYDS2011AAS...21813203S} mounted on the robotic 2m Faulkes Telescope North (FTN) at Haleakala, Hawaii, which is a part of the Las Cumbres Observatory. 
The wavelength coverage of FLOYDS is 3200--10000\,\AA{}. 
The spectra were taken using a 2\arcsec\, slit. Additional spectra were obtained with the Low Dispersion Survey Spectograph 3 (LDSS3; 4250--10000\,\AA{}) mounted on the 6.5m Magellan 2 - Clay Telescope at Las Campanas Observatory and the Binospec multislit imaging spectrograph (3850--9150\,\AA{}) located on the 6.5m Multiple Mirror Telescope (MMT) at the MMT Observatory (MMTO).

The FLOYDS spectra were reduced with the \texttt{floydsspec} pipeline\footnote{\url{https://github.com/svalenti/FLOYDS_pipeline/}}, following the methods described in \cite{2014MNRAS.438L.101V} to achieve flux and wavelength calibration, cosmic-ray removal, skyline correction, and for combining the red and blue orders per epoch. 

The LDSS3 and Binospec spectra were reduced using standard \texttt{Image Reduction and Analysis Facility} \citep[\texttt{IRAF};][]{iraf11986SPIE..627..733T,IRAF21993ASPC...52..173T} routines with the \texttt{twodspec} package. The spectra were bias-subtracted and flat-fielded, the sky background was modeled and subtracted from each image, and the one-dimensional spectra were optimally extracted. 
Wavelength and flux calibration were applied using arc lamps and standard stars taken near the time of observation. Telluric absorption was corrected with the \texttt{tellurics-begone.py}\footnote{\url{https://github.com/EJRidley/tellurics-begone}}\, script, which models and removes atmospheric $O_{2}$ and $H_{2}O$ features.

We obtained host-galaxy spectra via the DeVeny Optical Spectrograph mounted on the 4.3-meter Lowell Discovery Telescope \citep[LDT; ][]{LDT2012SPIE.8444E..19L,LDT2022SPIE12182E..27L} and the Alhambra Faint Object Spectrograph and Camera (ALFOSC) mounted on the 2.56m  Nordic Optical Telescope \citep[NOT;][PI: Blagorodnova]{NOT1_2010ASSP...14..211D} at the Spanish Roque de los Muchachos Observatory (ORM).

The DeVeny spectrum was taken with a spectral range of 3000--7400\,\AA{}, while the ALFOSC spectrum was taken through a 1\arcsec\ slit
with a spectral range of 3200--9600\,\AA{}. Both spectra were reduced using the \texttt{Python} package \texttt{PypeIt} \citep{pypeit:joss_pub}, following the standard procedure that includes bias and flat subtraction, flux calibration with a standard star and telluric correction.

We also retrieved the initial classification EFOSC2 spectrum \citep{EFOSC1984Msngr..38....9B} obtained with the 3.58 m ESO New Technology Telescope (NTT) at La Silla Observatory (between 3050 and 11000\,\AA{}), from the TNS \citep{TNS2021AAS...23742305G}, and a spectrum obtained with XSHOOTER \citep{2011A&A...536A.105V} on the ESO Very Large Telescope (VLT; PI: Lunnan) through the ESO archive. 
XSHOOTER is a multi-wavelength spectrograph covering 3000--25000\,\AA{}.
The XSHOOTER data were reduced and calibrated using the standard pipelines in \texttt{EsoReflex} \cite[v2.11.5;][]{EsoReflex}, and telluric absorption corrections were applied using \texttt{Molecfit} \cite[v1.5.9;][]{Molecfit1,Molecfit2}. Data in the wavelength range 6290--6384\,\AA{} were masked due to elevated noise.

Finally, we retrieved a pre-flare host-galaxy spectrum taken $\sim$28 days before the  discovery of AT\,2022csn as part of the main survey employed by the Dark Energy Spectroscopic Instrument (DESI) mounted on the 4-meter Mayall Telescope at Kitt Peak National Observatory (KPNO). DESI employs 5000 robotic fiber positioners covering 3600--9830\,\AA. We accessed the DESI Data Release 1 \citep[DR1;][]{desicollaboration2025datarelease1dark} spectrum via the Astro Data Lab \citep{data_lab_10.1117/12.2057445,astro_data_lab_NIKUTTA2020100411} and the SPectra Analysis and Retrievable Catalog Lab \citep[SPARCL;][]{SPARCL_juneau2025sparclspectraanalysisretrievable}.

We calibrate all of our spectra of AT\,2022csn to our extinction-corrected photometry using the \texttt{PySynphot} \citep{stsci2013ascl.soft03023S} package. The only exception is the second LDSS3 spectrum taken 273.5 days after the first $g$-band peak, as not enough photometric data is available for this time. The host-galaxy spectra were likewise calibrated to archival host photometry using \texttt{PySynphot}, except for the DESI spectrum, which was already spectrophotometrically flux calibrated. The DeVeny spectrum does not extend to the $i$-band wavelengths, preventing reliable flux calibration of its red end. It also does not cover the H$\alpha$ wavelength range. We therefore do not include it in our analysis. Our spectral observation log is presented in Table \ref{tbl:spectroscopic_log} and the spectra are shown in Figure \ref{fig:Stacked_spectra_norm}. All of our spectra will be uploaded to the Weizmann Interactive Supernova Data Repository \citep[WISeREP\footnote{\url{https://www.wiserep.org/}};][]{WISeREP}.

\begin{deluxetable*}{llllll}
\tablecaption{Log of spectroscopic observations.}
\tablehead{
\colhead{\shortstack[c]{\strut Phase \\
\strut (days)}} &
\colhead{\shortstack[c]{\strut Telescope/Instrument \\
\strut}} &
\colhead{\shortstack[c]{\strut Slit Width/Fiber diameter \\
\strut (\arcsec)}} &
\colhead{\shortstack[c]{\strut Exposure Time \\
\strut (s)}} &
\colhead{\shortstack[c]{\strut Grating/Grism \\
\strut}} &
\colhead{\shortstack[c]{\strut Resolution \\
\strut ($\lambda/\Delta\lambda$)}}
}
\startdata
-41.09\tablenotemark{a} & KPNO/DESI & 1.5 & 789 & B/R/Z & 2600/3650/4550\tablenotemark{b}\\
-8.86 & Las Cumbres/FLOYDS & 2 & 3600 & red/blu& 700/400\\
-7.12 & Las Cumbres/FLOYDS & 2 & 3600 &red/blu& 700/400\\
-2.76 & Las Cumbres/FLOYDS & 2 &3600& red/blu&700/400\\
-1.02 & NTT/EFOSC & 1 & 600& Gr\#13& 300\\
4.21 & Las Cumbres/FLOYDS & 2 & 3600& red/blu&700/400\\
8.56 & Las Cumbres/FLOYDS &  2 & 3600& red/blu& 700/400\\
12.92 & Las Cumbres/FLOYDS & 2 & 3600& red/blu&700/400\\
19.02 & VLT/XSHOOTER & 1.0/0.9 & 3600/3716& UVB/VIS& 5400/8900\\
23.37 & Magellan/LDSS3 & 0.75 & 1200 & VPH-ALL& 860\\
27.73& Las Cumbres/FLOYDS & 2 & 3600& red/blu& 700/400\\
36.44& Las Cumbres/FLOYDS &  2 & 1200& red/blu&700/400\\
199.33 & MMT/Binospec & 1 & 1200& G270& 1340 \\
273.37 & Magellan/LDSS3 & 0.75 & 1200& VPH-ALL& 860\\
963.27\tablenotemark{a} & LDT/DeVeny & 1 & 1800& DV2& 2300\\
969.36\tablenotemark{a} & NOT/ALFOSC & 1& 1350 & G4& 360
\enddata
\tablecomments{Phase is given in rest-frame days relative to the first $g$-band peak.}
\tablenotetext{a}{Host-galaxy spectrum}
\tablenotetext{b}{Average resolving powers derived from the published arm ranges: B (2000-3200), R (3200-4100) and Z (4100-5000).}

\end{deluxetable*}\label{tbl:spectroscopic_log}

\section{Analysis}\label{section:analysis}
\subsection{Peak Time}
One of the unusual properties of AT\,2022csn is the double peak in the light curve (seen most prominently in the $g$-band; Fig. \ref{fig:full_lightcurve}). We fit a second-degree polynomial around each of the \textit{g}-band peaks (between MJD 59638.5 and 59657.5 for the first peak and MJD 59658.5 and 59674.5 for the second peak). The best-fit $g$-band peak times are MJD 59648.17 $\pm$ 1.64 and 59666.47 $\pm$ 2.32, from which we deduce an 18.30 $\pm$ 2.84 day difference between the peaks.

\subsection{Blackbody Fits}

\begin{figure}[ht]
\includegraphics[width=0.4\textwidth]{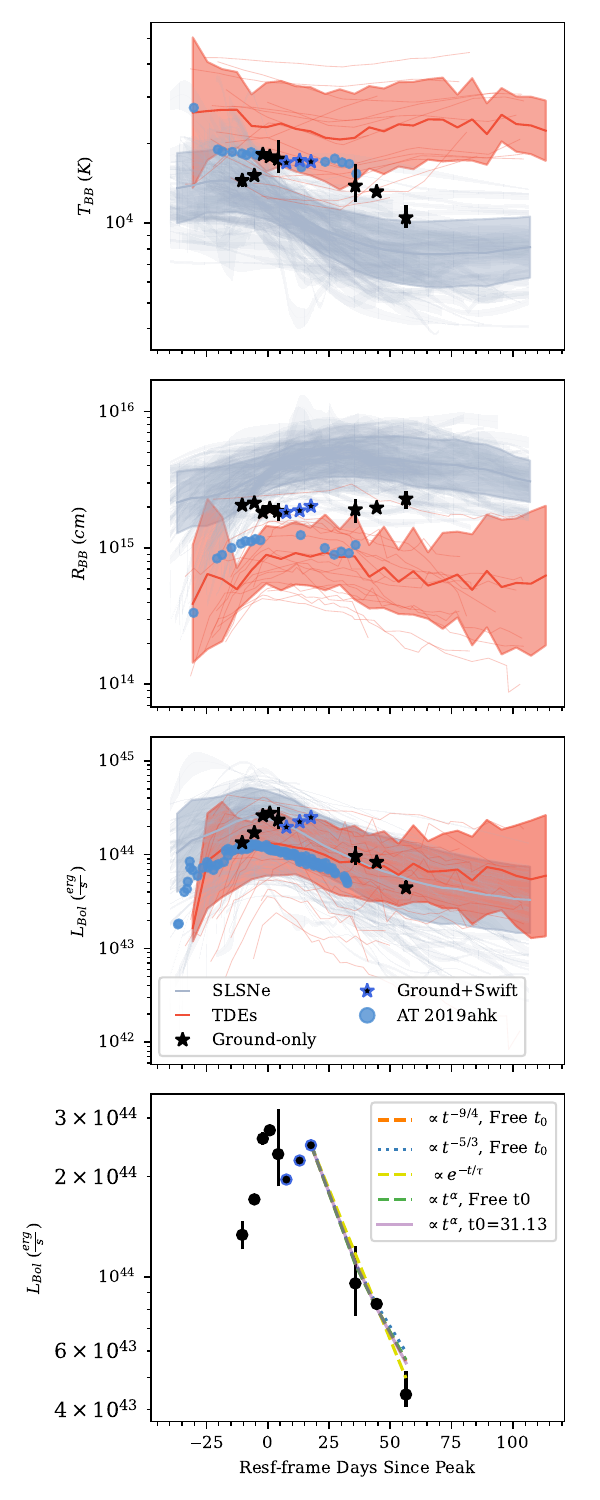}
\caption{From top: Best-fit blackbody temperature, radius, and resulting bolometric luminosity, and the bolometric luminosity decline fits of AT\,2022csn, compared with the TDE AT\,2019ahk \citep{Holoien_2019}, 17 TDEs from \cite{van_Velzen_2021} and 260 Type I SLSNe from \cite{gomez2024typeisuperluminoussupernova}. The mean and $1\sigma$ dispersion for each sample are shown as solid lines and shaded regions, respectively.
The photometric bands included in each epoch are listed in Table \ref{tbl:bolometric_epochs}. Error bars denote $1\sigma$ MCMC uncertainties.}
\label{fig:lightcurve_fitting}
\end{figure}

We fit a blackbody spectrum to the  photometry from ZTF ($gri$), Las Cumbres ($UBVgri$) and Swift ($uvw1\,uvw2\, uvm2$ and $ubv$)\footnote{The ATLAS $c$- and $o$-band photometry was not used in the blackbody fitting because their broad transmission curves overlap with the Las Cumbres $g$ and $r$ bands, so they do not provide independent measurements at the epochs where Las Cumbres data is also available.} using the \texttt{Light Curve Fitting}\footnote{\url{https://github.com/griffin-h/lightcurve_fitting}} package \citep{hosseinzadeh_2024_11405219}. 
We adopted the default MCMC configuration provided by the package, which uses 10 walkers, a minimum of three independent bands per epoch, and 200 burn-in steps followed by 100 sampling steps. We then use the best-fit blackbody parameters to calculate the bolometric luminosity using the Stefan-Boltzmann law. 
UV coverage is important for avoiding  systematic uncertainties when fitting hot blackbodies \citep{Arcavi_2022}. However, here we have few Swift UV epochs so we also fit epochs with ground-based data alone. The relatively low temperatures that we find for AT\,2022csn in the Swift epochs (see below) mitigate the problem of using ground-only data for the rest of the epochs.

\begin{deluxetable*}{llllll}
\tablecaption{Best-fit blackbody parameters and resulting bolometric luminosity.}
\tablehead{\colhead{Phase (days)} & \colhead{Bands} & \colhead{Temperature ($10^4$K)} & \colhead{Radius ($10^{15}$cm)} & \colhead{Luminosity ($10^{44}$ $\frac{\rm erg}{\rm s}$)}}
\startdata
$-10.435_{-0.031}^{+0.914}$  & $g r i$ & $1.48^{+0.13}_{-0.11}$ & $2.01_{-0.18}^{+0.20}$ & $1.37_{-0.15}^{+0.22}$ \\
$-5.472_{-1.466}^{+0.084}$  & $U B V g r i$ & $1.517^{+0.016}_{-0.015}$& $2.133_{-0.029}^{+0.030}$& $1.716_{-0.029}^{+0.026}$\\
$-2.0222_{-0.0109}^{+0.0078}$ & $UBVgri$ &$1.818^{+0.037}_{-0.046}$& $1.825_{-0.048}^{+0.042}$& $2.602_{-0.099}^{+0.111}$ \\
$0.807_{-0.034}^{+0.031}$ & $UBVgri$ & $1.767^{+0.049}_{-0.046}$ & $1.972_{-0.063}^{+0.036}$&$2.72_{-0.11}^{+0.10}$\\
$4.2313_{-0.0043}^{+0.0038}$ & $gri$ & $1.76^{+0.28}_{-0.19}$ & $1.85_{-0.23}^{+0.27}$& $2.34_{-0.40}^{+0.70}$\\
$7.60_{-0.22}^{+1.07}$ & \makecell[l]{$uvw2\,uvm2\,uvw1 \, ub$} & $1.696^{+0.035}_{-0.034}$ & $1.824_{-0.048}^{+0.052}$ & $1.961_{-0.061}^{+0.056}$\\
& $UBVgri$ &  & & \\
$12.989_{-1.004}^{+0.056}$ & \makecell[l]{$uvw2\, uvm2\,uvw1\,ubv$} & $1.727^{+0.024}_{-0.027}$ & $1.878_{-0.032}^{+0.035}$ & $2.238_{-0.063}^{+0.054}$\\
& \makecell[l]{$UBVgri$} & & & \\
$17.614_{-1.092}^{+0.021}$ & \makecell[l]{$uvw2\, uvm2\,uvw1\,ubv$} & $1.709^{+0.029}_{-0.030}$ & $2.025_{-0.041}^{+0.044}$ & $2.487_{-0.067}^{+0.073}$ \\
& \makecell[l]{$UBVgri$} & & & \\
$35.7292_{-0.0211}^{+0.0084}$&  $BVgr$ & $1.49^{+0.40}_{-0.27}$& $1.74_{-0.43}^{+0.46}$ & $1.05_{-0.24}^{+0.52}$ \\
$44.4275_{-0.0124}^{+0.0092}$ & $UBVgri$ & $1.332^{+0.032}_{-0.035}$ & $1.935_{-0.066}^{+0.083}$ & $0.841_{-0.023}^{+0.023}$\\
$56.3833_{-0.0039}^{+0.0031}$& $gri$ & $1.17^{+0.25}_{-0.19}$ & $1.94_{-0.41}^{+0.54}$ & $0.52_{-0.10}^{+0.18}$
\enddata
\tablecomments{Phase is given in rest-frame days relative to the first $g$-band peak with upper and lower errors denoting the binning epochs.}
\end{deluxetable*}\label{tbl:bolometric_epochs}

Our results are shown in Table \ref{tbl:bolometric_epochs} and Figure \ref{fig:lightcurve_fitting}, where we also compare AT\,2022csn to other TDEs (including AT\,2019ahk which was in a similar host galaxy as AT\,2022csn) and to Type I SLSNe. The blackbody temperature of AT\,2022csn lies at the lower end of the TDE sample, while the blackbody radius is located at the high end of the TDE sample. In both parameters, AT\,2022csn borders on the region characteristic of Type I SLSNe. AT\,2022csn is on the high end of the bolometric luminosity for TDEs, especially around its peaks. 

We further fit the post-peak (i.e. after the second $g$-band maximum) bolometric light curve between MJD 59681 and 59857 using both power-law ($L \propto \left(\frac{t - t_0}{\tau}\right)^{-\alpha}$) and exponential ($L \propto e^{-t / \tau}$) declines. For the power-law decline we test the canonical full disruption with fixed $\alpha = 5/3$, the partial disruption fixed $\alpha = 9/4$ \citep{Coughlin_2019,Ryu_2020_pTDE}, a fit with free $\alpha$ and $t_0$ fixed at $31.13$ days (based on the disruption date from the best-fit TDE model described in Section \ref{section:tde models}), and with both $\alpha$ and $t_0$ as free parameters. The best-fit parameters and reduces chi squared values are listed in Table \ref{tbl:decline rates} (for the case with both $\alpha$ and $t_0$ free, the luminosity normalization factor had to be fixed to ensure convergence, and for the canonical $\alpha = 5/3$ model, we required $t_0$ to occur before the discovery epoch). The limited number post-peak epochs prevents a definitive conclusion regarding the preferred model.
\begin{deluxetable*}{lll}
\tablecaption{Best-fit bolometric decline parameters.}
\tablehead{\colhead{Model} & \colhead{Parameters} & \colhead{$\chi^2_{red}$}}
\startdata
$\alpha=5/3$, Free $t_0$ & $t_0=$ 10.54 $\pm$ 1.41 & 7.41\\
$\alpha=9/4$, Free $t_0$ & $t_0=$ 23.57 $\pm$ 4.79 & 5.16 \\
Free $\alpha$, $t_0=$ 31.13 & $\alpha=$ 2.58 $\pm$ 0.36& 4.68 \\
Free $\alpha$ and $t_0$ & $t_0=19.34 \pm 1.38$ & 2.74\\
&  $\alpha =$ 2.071$\pm$ 0.017& \\
Exponent Free $\tau$ & $\tau =$ 24.07 $\pm$ 0.88 & 1.91
\enddata
\tablecomments{$t_0$ and $\tau$ are given in days.}
\end{deluxetable*}\label{tbl:decline rates}

\subsection{TDE Model Fits}\label{section:tde models}
 
We fit the \cite{Guillochon_2013} and \cite{Mockler2019ApJ...872..151M} reprocessing-emission model implemented in  the Modular Open Source Fitter for Transients \citep[\texttt{MOSFiT};][]{MOSFiT2018ApJS..236....6G} and the \cite{Piran_2015} stream-collision model implemented in the \texttt{TDEMass} package \citep{Ryu_2020} to our data.

The reprocessing-emission model converts the post-disruption mass fallback rate onto the SMBH directly into bolometric luminosity through a constant radiative efficiency parameter, $\epsilon$ and the viscous timescale, $T_{\rm viscous}$ \citep{Mockler2019ApJ...872..151M}, calibrated to hydrodynamical simulations of TDEs of polytropic stars \citep{Guillochon_2013}. The radiation is then modeled as a blackbody photosphere of radius $R_{\rm phot}$. The free parameters are the SMBH mass ($M_{\rm BH}$), the stellar mass ($M_\ast$), $T_{\rm viscous}$, $\epsilon$, $R_{\rm phot} \propto R_{\rm ph,0} \times L^l$ (with $R_{\rm ph,0}$ and $l$ as free parameters), the scaled impact parameter ($b$, a proxy for $\beta \equiv R_t/R_p$, with $R_t$ the tidal radius and $R_p$ the pericenter of the original stellar orbit), the time of first fallback ($t_{\rm exp}$, denoted $t_0$ in the decline fits), the host-galaxy column density ($n_H$), and a white-noise term ($\sigma$). As noted by \cite{Mockler2019ApJ...872..151M} and \cite{Mockler_Remirez2021ApJ...906..101M}, this model has several limitations. For example, the model relies on simplifications of the complex physics involved and the stars are assumed to be zero-age main sequence solar metallicity stars with a blend of 4/3 and 5/3 polytropes. In addition there is a strong degeneracy
between the stellar mass that falls onto the SMBH and the efficiency of the radiative process. 

\begin{figure*}[ht]
\includegraphics[width=\textwidth]{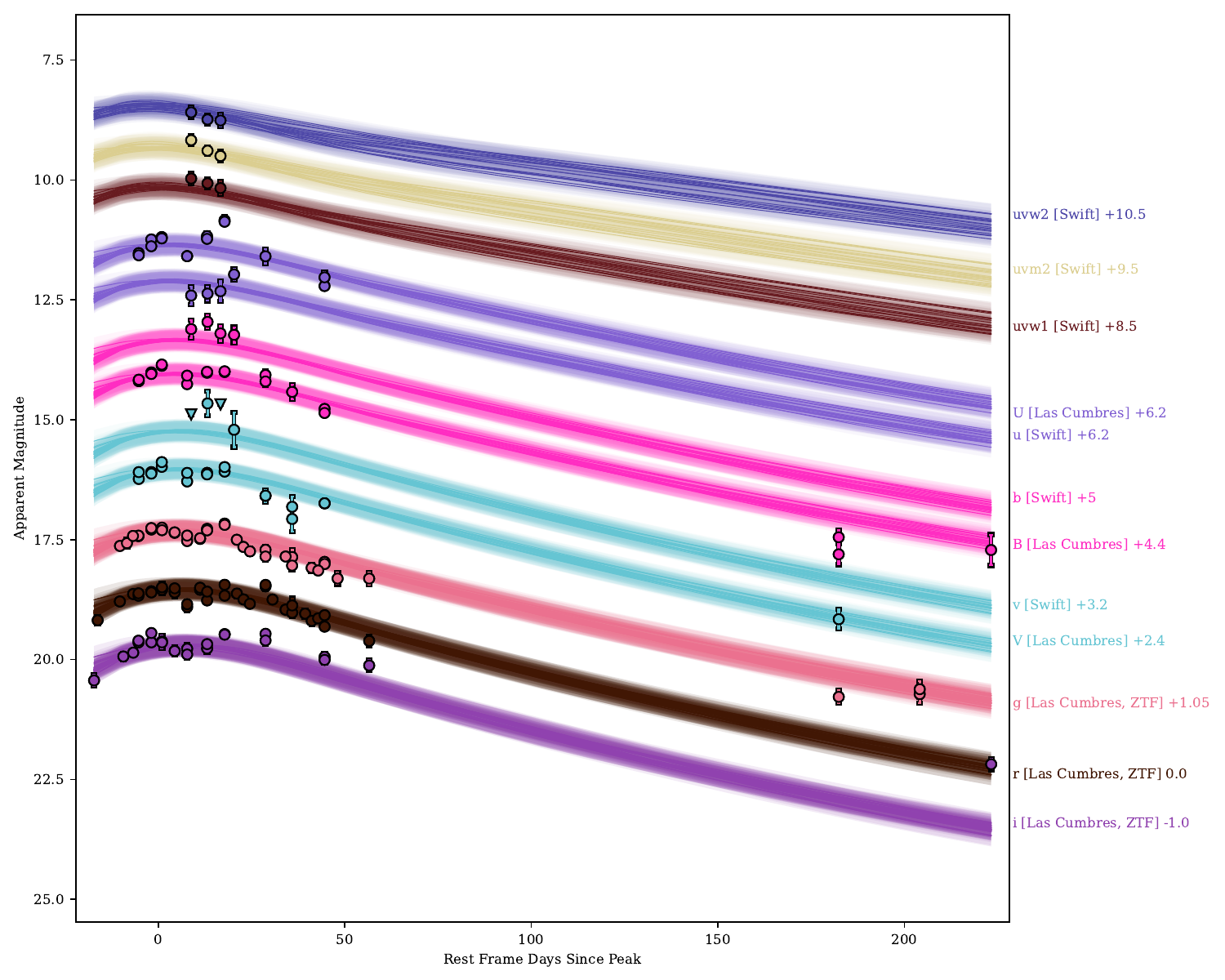}
\caption{Nested sampling fits from MOSFiT to the photometry of AT\,2022csn. The solid lines represent the median light curves derived from the ensemble of model samples, while the shaded regions illustrate the spread (variance) within those samples. Arrows indicate $3\sigma$ nondetection upper limits. Bands and offsets are indicated on the right side of the figure.}
\label{fig:mosfit_lightcurve}
\end{figure*}

We perform the fit using the nested sampling algorithm implemented in \texttt{DYNESTY} \citep{Speagle_2020}.
Consistent with the blackbody fits, we exclude ATLAS bands and adopt the \texttt{MOSFiT} default termination criterion based on an expected remaining evidence gain of 0.02. We adopt the default priors of \cite{Mockler2019ApJ...872..151M} for the fit. The resulting best-fit parameters are listed in Table \ref{tbl:mosfit params} (including systematic uncertainties arising from model simplifications as quantified by \citealt{Mockler2019ApJ...872..151M}), the fit is shown in Figure \ref{fig:mosfit_lightcurve} and the  posterior distributions are shown in Figure \ref{fig:mosfit_corner} in Appendix \ref{appendix:MOsfit corner}. The model can reproduce the overall shape of the light curve, but not its double peak.

\begin{deluxetable*}{llll}
\tablecaption{Best-fit reprocessed-emission MOSFiT parameters with $1\sigma$ confidence intervals.}
\tablehead{\colhead{Parameter} & \colhead{Best-Fit Value} & \colhead{Systematic Error} & \colhead{Units}}
\startdata
log $(R_{ph,0})$ & $1.51^{+0.13}_{-0.13}$ & $\pm$0.40 & ... \\
log $(T_{viscous})$ & $-0.5^{+0.92}_{-1.43}$ & $\pm$0.10  & days \\
b  & $0.52^{+0.05}_{-0.06}$ & $\pm$0.35  & ... \\
log $(M_{BH})$ & $6.71^{+0.07}_{-0.06}$ & $\pm$0.20  & $M_\odot$ \\
log $(\epsilon)$ & $-2.03^{+0.15}_{-0.11}$ & $\pm$0.68  & ... \\
l & $0.97^{+0.08}_{-0.07}$ & $\pm$0.20  & ... \\
log $(n_H)$ & $18.84^{+0.8}_{-1.13}$ & $...$  & cm$^{-2}$ \\
$M_\ast$ & $1.00^{+0.03}_{-0.03}$ & $\pm$0.66  & $M_\odot$ \\
$t_{exp}$ & $-31.13^{+2.21}_{-2.83}$ & $\pm$15.0  & days \\
log $(\sigma)$ & $-0.71^{+0.03}_{-0.03}$ & ...  & ... \\
\enddata
\tablecomments{The systematic errors are the MOSFiT systematic uncertainties associated with simplifying assumptions in the model estimated by \cite{Mockler2019ApJ...872..151M}.}
\end{deluxetable*}\label{tbl:mosfit params}

For the stream-collision model, \texttt{TDEMass} estimates the stellar and SMBH masses by solving two coupled nonlinear equations \citep[Equations 9 and 10 of ][]{Ryu_2020} that relate these quantities to the observed peak bolometric luminosity and the temperature at peak. The equations depend on the function $\Xi(M_{\rm BH}, M_\star)$, which describes the width of the debris energy distribution, as well as two weakly constrained physical parameters: $c_1$, which sets the characteristic radius where shock energy is dissipated, and $\Delta\Omega$, the solid angle of the emitting region (in units of $\pi$). Because these quantities are uncertain, the default values $c_1 = 1$ and $\Delta\Omega = 2$ are generally adopted. The updated version of the code can also include cooling effects following \citet{krolik2025followmassconcordance}, in which case the masses are obtained by solving their Equations 17 and 19. 
We use the bolometric luminosity and corresponding temperature at the second $g$-band peak, which unlike the first peak includes both Swift and ground-based measurements, allowing for a more reliable temperature estimate \citep{Arcavi_2022}.
The combination of high peak luminosity ($2.487^{+0.073}_{-0.067} \times 10^{44}\,{\rm erg\,s^{-1}}$), and low temperature at the second $g$-band peak ($17{,}089^{+291}_{-302}\,{\rm K}$), drives the fit towards high-mass stars, finding:
$$\log_{10}\left( \frac{M_{BH}}{M_\odot} \right) = 6.826_{-0.0116}^{+0.0097},  \quad M_*= 14.00_{-1.20}^{+1.60}\ {M_\odot}$$
When incorporating the cooling effects of \citet{krolik2025followmassconcordance}, the best fit becomes:
$$ \log_{10}\left( \frac{M_{BH}}{M_\odot} \right) = 7.079_{-0.016}^{+0.013} ,\quad M_*=6.80_{-0.74}^{+0.91}\ M_\odot $$

\subsection{Spectroscopy}\label{section:analysis_spectroscopy}
\subsubsection{Emission Line Analysis}
\begin{figure}[ht]
\centering    \includegraphics[width=0.5\textwidth]{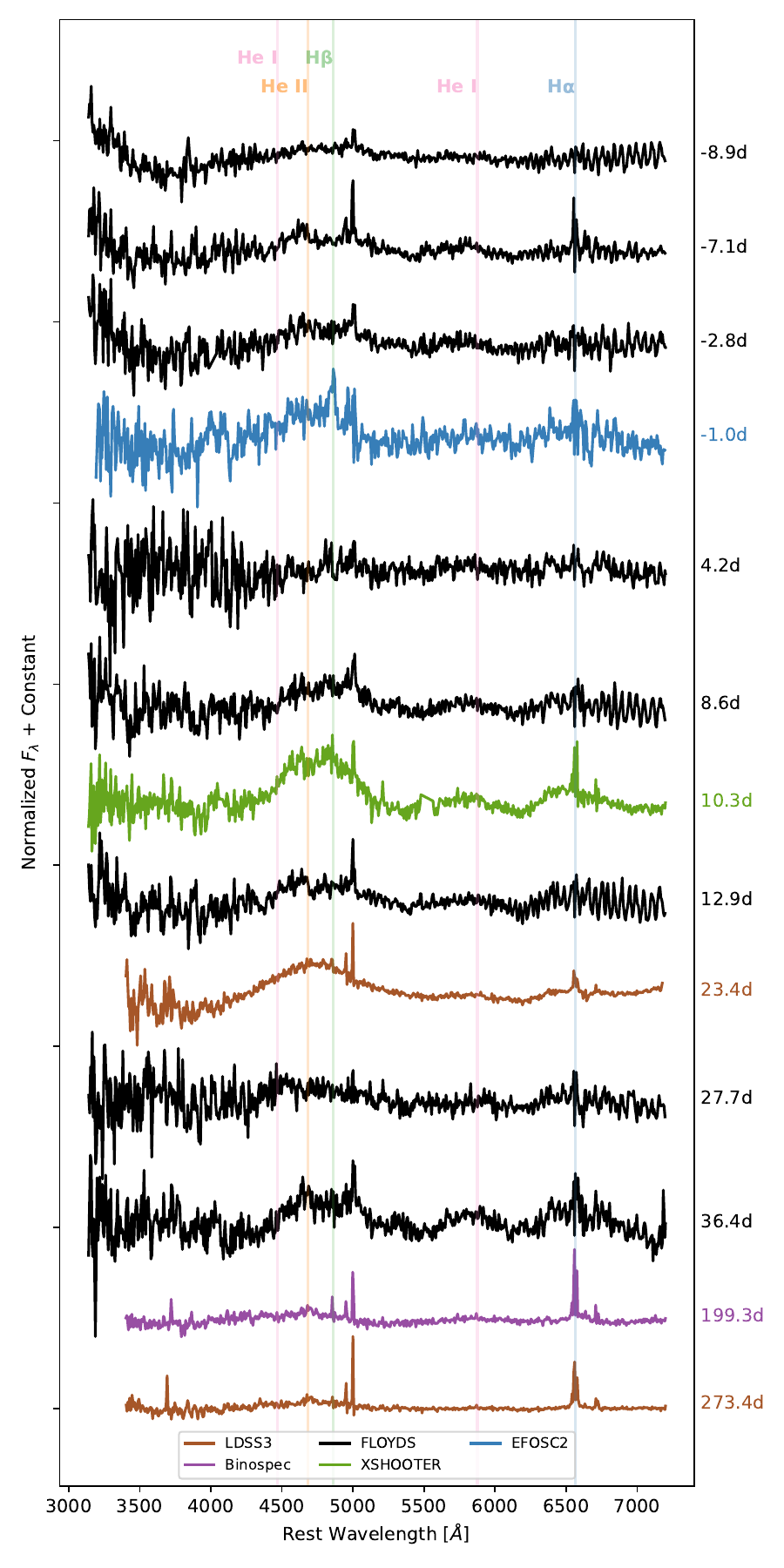}
\caption{Host-galaxy and continuum subtracted spectra of AT\,2022csn. Rest-frame days relative to the first $g$-band peak are noted.}
\label{fig:Stacked_spectra_host_cont}
\end{figure}

We follow the procedure described by \citet{charalampopoulos2022detailed} to remove  host-galaxy and continuum components from the TDE spectra (i.e., up to 273.5 rest-frame days post first $g$-band peak). For the host-galaxy subtraction, we used the DESI spectrum obtained $\sim$28 days prior to discovery of the TDE. The spectrum shows no evidence of TDE features and provides the widest wavelength coverage and highest resolution among the available host-galaxy spectra. Since the DESI spectrum has higher spectral resolution than all of our TDE spectra (except for the spectrum obtained with XSHOOTER), we resampled it to the wavelength values of each TDE spectrum individually. For the XSHOOTER data, we instead resampled the XSHOOTER spectrum to match the DESI wavelength values (resampling was performed using the \texttt{Scipy interp1d}\footnote{\url{https://docs.scipy.org/doc/scipy/reference/generated/scipy.interpolate.interp1d.html}} function).
Variations in seeing conditions, slit widths, position angles, and the fixed fiber aperture used for the DESI observation resulted in different host-galaxy contributions to each of the TDE spectra.
To account for this, we subtracted different scalings of the host flux from each spectrum, adjusting the scaling based on the strength of the narrow [\ion{O}{3}] $\lambda5007$ and H$\alpha$ emission lines. While this approach minimizes host-galaxy contamination, it is not perfect, and galaxy residuals may still remain. Our host-subtracted spectra are presented in Figure \ref{fig:Stacked_spectra_host} in Appendix \ref{appendix:Stacked_spectra_host}.
Next, we fit and remove the spectral continuum using a third-order polynomial fit to line-free regions. We adopt common TDE line-free wavelength ranges \citep[namely, 3900--4000\,\AA{}, 4220--4280\,\AA{}, 5100--5550\,\AA{}, 6000--6350\,\AA{}, and 6800--7000\,\AA{};][]{charalampopoulos2022detailed} together with an additional line-free region (3350-3420\,\AA{}) customized to the AT\,2022csn spectra. Our host-galaxy and continuum subtracted spectra are presented in Figure \ref{fig:Stacked_spectra_host_cont}.

\begin{figure}[ht]
\centering    \includegraphics[width=0.5\textwidth]{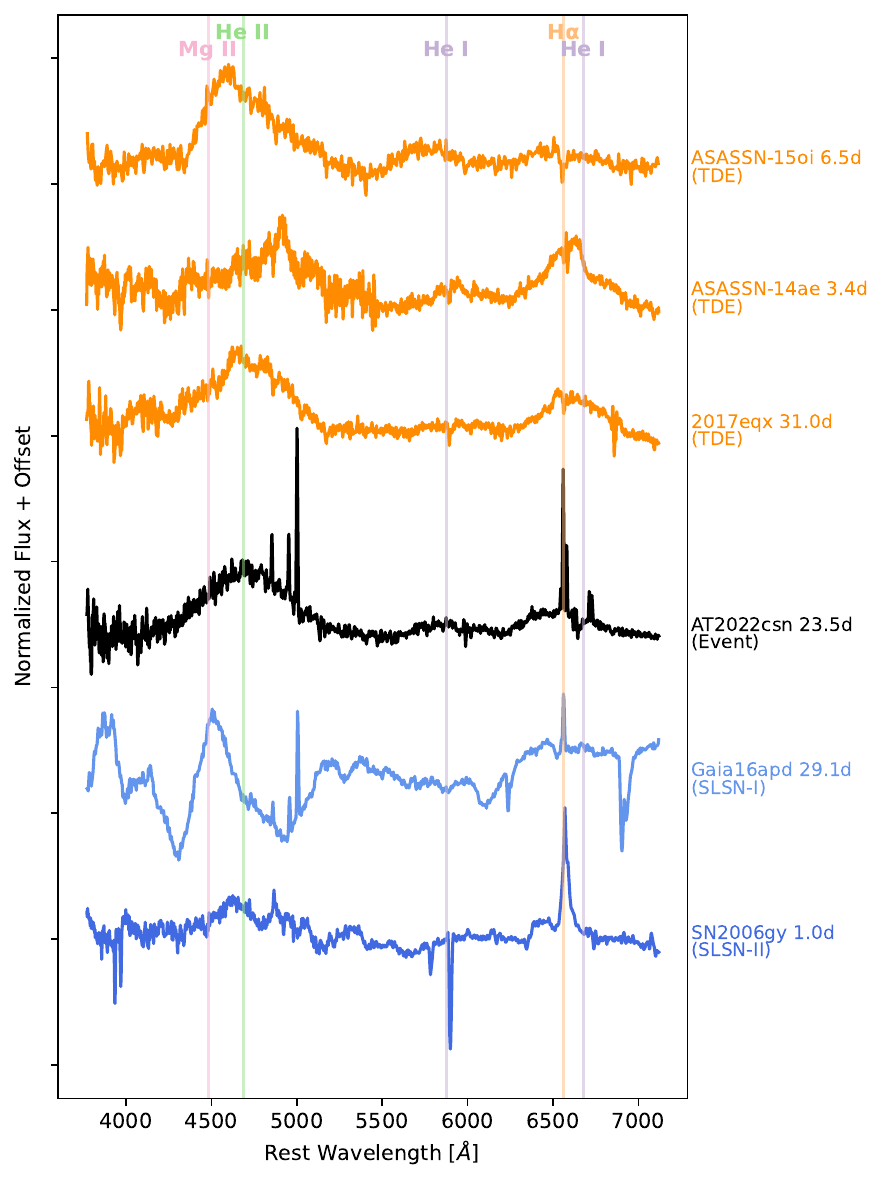}
\caption{Comparison between the spectral features of AT\,2022csn to those of various H+He TDEs at similar epochs and to those of SLSNe.
\label{fig:He_II_stacked}}
\end{figure}

The host-galaxy and continuum subtracted spectra of AT\,2022csn show clear broad \ion{He}{2} $\lambda4686$, H$\alpha$, and \ion{He}{1} $\lambda5876$ emission features. The \ion{He}{2} and \ion{He}{1} features are prominent already 2.8 days before the first $g$-band peak, while H$\alpha$ becomes clearly visible only at 8.6 days post peak. Most broad emission features fade at late times (199.3 and 273.4 days after the first $g$-band peak), although \ion{He}{2} remains detectable. 
Narrow lines (such as those of [\ion{O}{3}], H$\beta$, H$\alpha,$ [\ion{N}{2}]) are likely from residual host-galaxy emission (a detailed analysis of the host-galaxy emission lines is presented in Section \ref{section:host galaxy properties}).

The broad features seen in the spectra are consistent with those seen in other optical/UV TDEs, as shown in Figure \ref{fig:He_II_stacked}, where we compare the LDSS3 spectrum of AT\,2022csn taken 23.5 days post first $g$-band peak, to spectra of the H+He TDEs ASASSN\,14ae \citep{Holoien_2014}, ASASSN\,15oi \citep{Holoien_2016} and AT\,2017eqx \citep{AT2017eqx} at similar epochs, retrieved via WISeREP \citep{WISeREP}. These features are not consistent with the spectral signatures of Type I and Type II SLSNe (Fig. \ref{fig:He_II_stacked}), such as the \ion{O}{2} absorption complex seen in Gaia16apd and other Type I SLSNe \citep[e.g.][]{Kangas_2017,Quimby_2018,Gal_Yam_2019} and the narrow, interaction driven hydrogen emission or P-Cygni absorption Balmer features exhibited in SN\,2006gy and other Type II SLSNe  
\citep[e.g.][]{2007ApJ...666.1116S,Gal_Yam_2019_The_Most_Luminous_SNe}.

\section{Host Galaxy Analysis}\label{section:host_galaxy}

\subsection{Host Galaxy Properties}\label{section:host galaxy properties}
\subsubsection{Archival Photometry} 
We checked the ZTF and ATLAS forced photometry for historic variability in the optical light curve (Figure \ref{fig:archival_lightcurve} in Appendix \ref{appendix:ZTF ATLAS archival}).
We find no significant optical variability prior to or after the TDE at the $5\sigma$ level. 
The NEOWISE $W1$ and $W2$ data shown in Figure \ref{fig:NEOWISE} does show some variability (though unlike the ZTF and ATLAS search, these are unsubtracted magnitudes). Using the $W1-W2$ color diagnostic for AGN of $>$0.7 \citep{Hickox_2017} and $>$0.8 magnitudes \citep{NEOWISE_0_8_2012ApJ...753...30S}, we find that most epochs lie above the more inclusive threshold, with several also exceeding the stricter one (although they are within $\sim1\sigma$ of these selection boundaries) indicating elevated MIR emission marginally consistent with an AGN.

\subsubsection{Host Galaxy SED Modeling}\label{section:blast}
We extract best-fit host galaxy physical parameters from \texttt{Blast}\footnote{\url{https://blast.scimma.org/}}
\citep{jones2024blastwebapplicationcharacterizing} for the host galaxy of AT\,2022csn. \texttt{Blast} associates transients with likely host galaxies and performs host-galaxy SED fitting to archival imaging using the \texttt{Prospector} framework \citep{Leja_2017,Johnson_2021}. The parameters include the galaxy mass, star formation rate, metallicity, age, AGN fraction, and star formation history (the full list of the 14 parameters is provided in Appendix \ref{appendix:Blast_parameters}). We also retrieve the best-fit \texttt{Blast} parameters for the available 26 hosts out of the 33 TDEs in \citet[][the fit failed for the remaining events due to insufficient data]{van_Velzen_2020}, as well as for AT\,2019ahk, for comparison.

\begin{figure*}[ht]
\includegraphics[width=\textwidth]{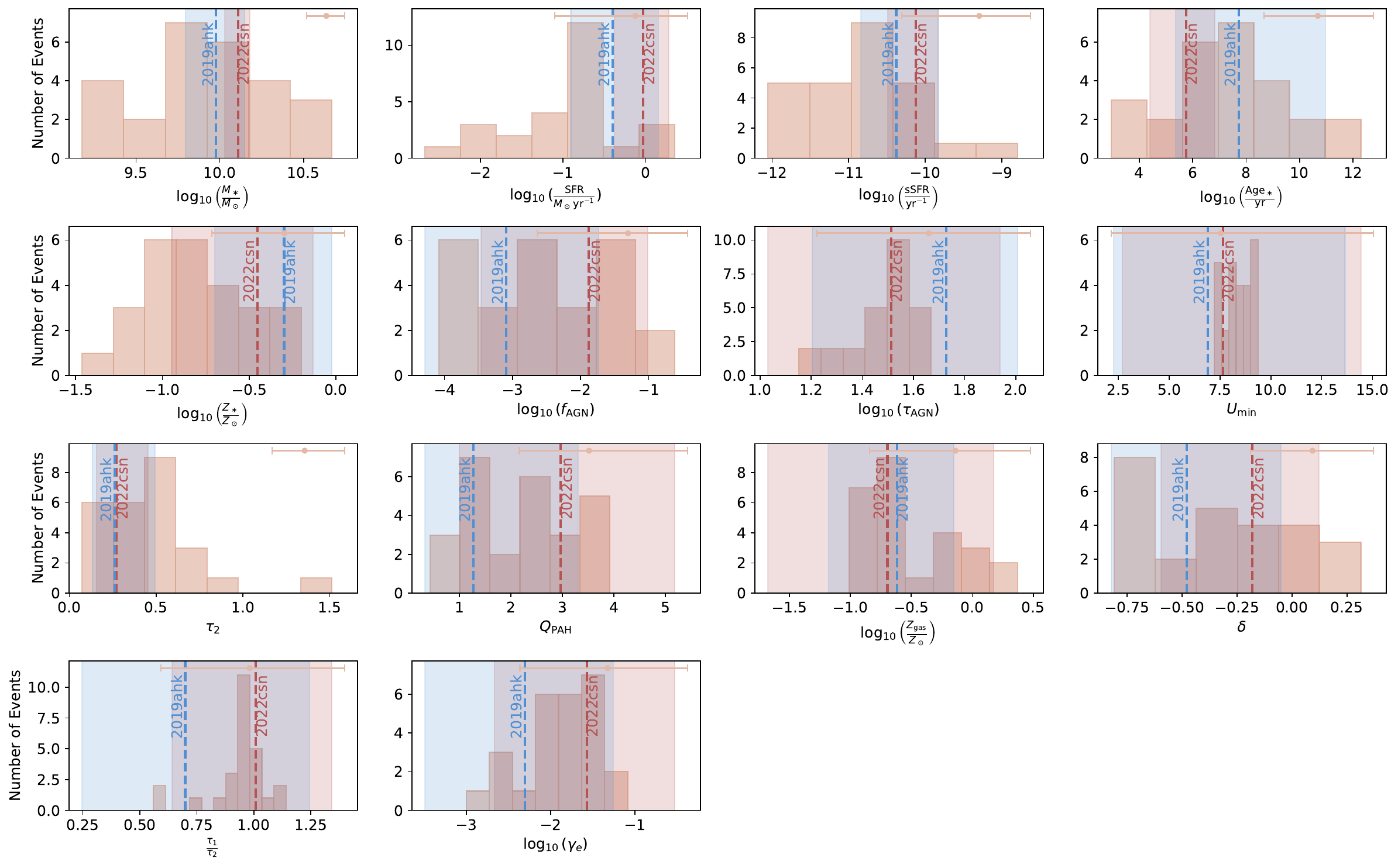}
\caption{\texttt{Blast} best-fit host galaxy parameters for the hosts of AT\,2022csn (the red dashed line and red shaded region denote the median and 1$\sigma$ posterior distribution width respectively), AT\,2019ahk (blue) and the available 26 TDE hosts from the \citet{van_Velzen_2020} TDE sample (orange). The average 1$\sigma$ upper and lower posterior distribution widths of each parameter from the comparison sample are marked in orange at the upper right corner of each panel. }
\label{fig:Blast_hist}
\end{figure*}

The results are plotted in Figure \ref{fig:Blast_hist}. Across all parameters, the host galaxies of AT\,2022csn and AT\,2019ahk are consistent with the distribution of the comparison TDE host-galaxy sample within the uncertainties.

\subsubsection{Narrow Emission Lines}\label{section:BPT}
\begin{figure}[ht]
\centering    \includegraphics[width=0.5\textwidth]{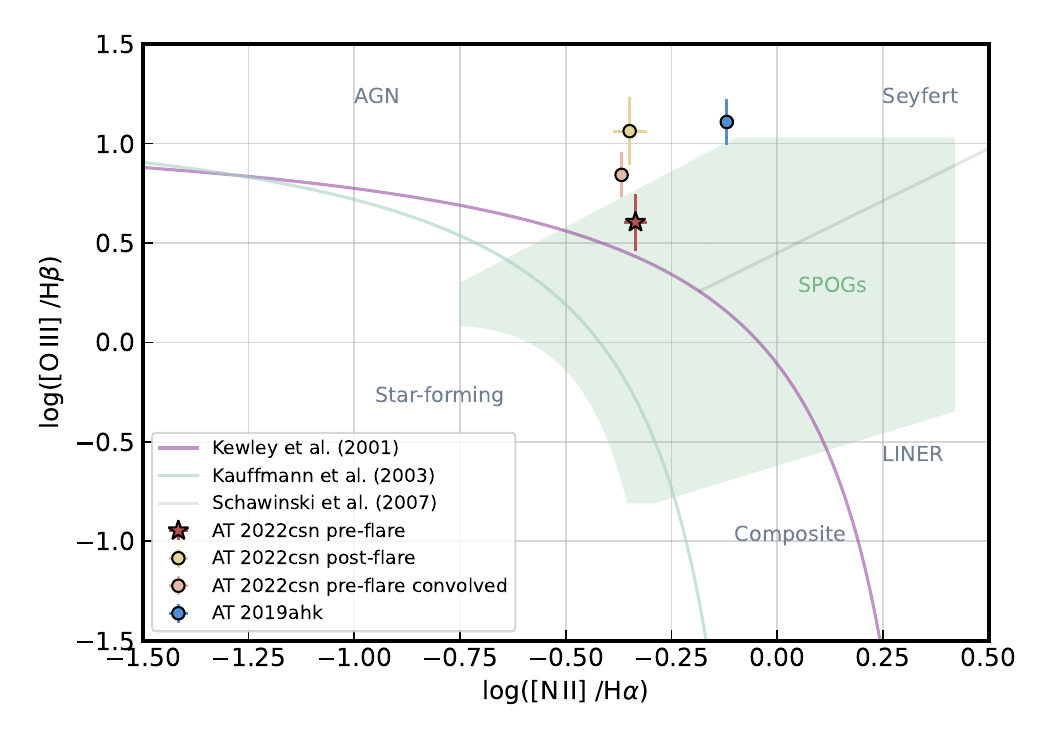}
\caption{BPT diagram for the host galaxy of AT\,2022csn using pre-flare, post-flare and pre-flare convolved to post-flare resolution spectra, together with the host galaxy of AT\,2019ahk \citep{Holoien_2019}.
Regions defined using the theoretical maximum starburst from \citet{Kewley_2001}, the observed starburst-AGN separation from \citet{Kauffman_2003}, the division between Seyfert and LINER galaxies according to \citet{schawinski_2007} and the population of Shocked Post-Starburst (SPOGs) Galaxies from \cite{Altalo_2016ApJS..224...38A} are also shown.}
\label{fig:BPT}
\end{figure}

We use the \texttt{PyQSOFit}\footnote{\url{https://github.com/legolason/PyQSOFit}} package \citep{pyqsofit_1_2018ascl.soft09008G,pyqsofit_2_2019ApJS..241...34S}, to measure the flux in the narrow emission lines seen in the DESI pre-flare host galaxy spectrum. The fit is presented in Figure \ref{fig:PyQSOFit_DESI} in Appendix \ref{appendix:host_galaxy_spectral_fits}. We find  $\log_{10}\left(\frac{[N_{II}]}{H\alpha}\right) =$ -0.335 $\pm$  0.026 and $\log_{10}\left(\frac{[O_{III}]}{H\beta}\right) =$ 0.60 $\pm$ 0.14. These ratios place the host galaxy in the AGN-Seyfert region of the Baldwin, Phillips \& Terlevich  \citep[BPT;][]{BPT_1981PASP...93....5B} diagram (Fig. \ref{fig:BPT}), as well as in the area associated with Shocked Post Starburst Galaxies \citep[SPOGs;][see below]{Altalo_2016ApJS..224...38A}.

Repeating this analysis for the post-flare ALFOSC spectrum yields slightly different line ratios, but the host galaxy remains firmly within the AGN-Seyfert region (Fig. \ref{fig:BPT}). To check whether the small difference in line ratios can be explained by the differences in spectral resolution between the DESI and ALFOSC spectra, we convolved the DESI spectrum with the ALFOSC resolution using a non-stationary Gaussian kernel and repeated the analysis, with uncertainties estimated via Monte Carlo sampling. The convolved DESI spectrum line ratios shift towards the ALFOSC ones, indicating that indeed the offset can most likely be explained by differences in the spectral resolution between the two instruments.

We also check the [\ion{O}{3}]/H$\beta$ vs. [\ion{S}{2}] $\lambda\lambda$6716,6731/H$\alpha$ line ratios of the DESI spectrum and find that the host galaxy is consistent with an AGN-Seyfert in this phase space as well (Figure \ref{fig:BPT_SII} in Appendix \ref{appendix:host_galaxy_spectral_fits}). 
Overall, these results consistently suggest the presence of AGN-Seyfert activity both before and after the TDE. 

\begin{figure}[ht]
\centering    \includegraphics[width=0.5\textwidth]{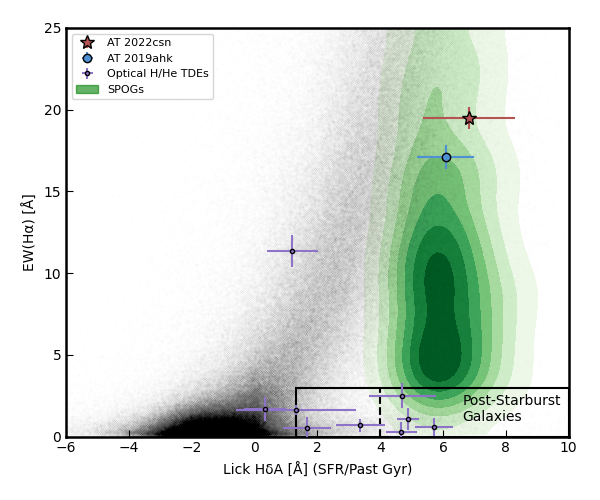}
\caption{H$\alpha$ EW vs. Lick H$\delta_A$ index of the host galaxies of AT\,2022csn (red), AT\,2019ahk \citep{Holoien_2019}, H/He TDEs \citep{French_2016} and SDSS galaxies \citep[gray;][]{Strauss_2002}. Regions associated with SPOGs \citep{Altalo_2016ApJS..224...38A} and with post-starburst and quiescent Balmer strong galaxies \citep[dashed and solid lines, respectively;][]{French_2016} are also shown.}
\label{fig:EW_Lick}
\end{figure}

We next measure the H$\alpha$ equivalent width (EW) and the Lick H$\delta_A$ absorption index \citep{Worthey_1997} by fitting the local continuum and integrating the emission or absorption relative to it, respectively. We find an H$\alpha$ EW of 19.47 $\pm$ 0.67 \AA{} and H$\delta_A$ = 6.81 $\pm$ 1.4 \AA{}. 
We plot these results in Figure \ref{fig:EW_Lick} together with the host galaxy of AT\,2019ahk \citep{Holoien_2019}, the sample of H/He TDE host galaxies used in \cite{French_2016}, galaxies from the SDSS main spectroscopic survey \citep{Strauss_2002}, and the region associated with SPOGs from \cite{Altalo_2016ApJS..224...38A}\footnote{\url{https://spogs.org/}}. SPOGs are characterized by strong Balmer absorption from A Type stars, indicative of substantial star formation within the past $\sim$Gyr, combined with emission-line properties consistent with shocks or AGN activity, rather than ongoing star formation.
We also mark the region of phase space associated with post-starburst and quiescent balmer-strong galaxies following \citet{French_2016,French_2020}. These regions have similar Balmer absorption from A Type stars, but show very little or no H$\alpha$ emission. They include only 0.2\% and 2.3\% of the
SDSS galaxies \citep{Strauss_2002}, yet encompass 38\% and 75\% of optical/UV TDE host galaxies, respectively \citep{French_2016}. 

While the H$\delta_A$ absorption of the host of AT\,2022csn is comparable to those of other TDE hosts, it shows significantly stronger H$\alpha$ emission, consistent with the area associated with SPOGs. Most SPOGs exhibit star-formation histories similar to those of traditionally selected post-starburst galaxies, but are, on average, younger and may therefore experience higher levels of dust obscuration \citep{French_2018}.

\subsection{Black Hole Mass Estimate}\label{section:galaxy mass estimates}
\subsubsection{M-$\sigma_\ast$ Relation}
We measure the stellar velocity dispersion ($\sigma_\ast$) from the host-galaxy spectrum taken with DESI using the \texttt{pPXF} penalized pixel-fitting code \citep{cappellari_2023MNRAS.526.3273C}. This method fits the observed spectrum with a convolution of stellar population templates and a line-of-sight velocity distribution (LOSVD), optimizing the fit via a maximum penalized likelihood approach while accounting for the instrumental spectral resolution. For our template library, we adopted stellar population models based on the high-resolution ($R \sim 10{,}000$) XSHOOTER Spectral Library \citep{Verro_2022}. 
 
We perform the fit over the wavelength range 3880--5550\,\AA{} (where the DESI instrumental dispersion is $\sim$ 38\,km/s), which includes the \ion{Ca}{2} H+K $\lambda$3969, $\lambda$3934 and \ion{Mg}{1} $\lambda$5175 absorption features commonly used for measuring $\sigma_\ast$ \citep{Koss_2022}.
The model includes fourth-degree additive Legendre polynomials to mitigate template mismatch and correct low-order continuum differences. Narrow spectral regions were masked to exclude emission line contamination following the ``strong" and ``weak" masking windows defined by \citet{Koss_2022} with additional masking of H$\alpha$. The LOSVD is then described using the first four Gauss-Hermite moments. The spectrum best-fit model and residuals are presented in Figure \ref{fig:ppxf} in Appendix \ref{appendix:host_galaxy_spectral_fits}.
To asses uncertainties we use both the formal \texttt{pPXF} covariance-matrix errors and an independent Monte Carlo approach, and adopt the larger uncertainty of the two.

We find a $\sigma_\ast$ value of 59.84 $\pm$ 27.11\,km/s.
Using the $M$-$\sigma_\ast$ relation of \citet{Kormendy_2013}, we obtain:
$$\log_{10} \left(\frac{M_{BH}}{M_\odot} \right) = 6.18 \pm 0.93$$ which is consistent with the results obtained from modeling the TDE emission with both \texttt{MOSFiT} and \texttt{TDEmass}.

\subsubsection{Bulge-SMBH Mass Relation}

An alternative SMBH mass estimate can be obtained using the bulge-SMBH mass relation \citep[e.g.,][]{Kormendy_2013}. Following \citet{Wevers} and \citet{ramsden2025evidencesteepersmbhbulgemass}, we estimate the bulge-to-total ($B/T$) light fraction in the $g$-band by adopting the Pan-STARRS DR1 \citep{PS1_2017yCat.2349....0C} PSF magnitude as a proxy for the bulge component (as it traces the centrally concentrated light approximated by a point source), and the Kron magnitude as a measure of the total galaxy light. We find $B/T=0.767\pm0.016$. 

We alternatively estimate the $B/T$ ratio using the average stellar mass to bulge mass ratio inferred for several TDE host galaxies \citep{Holoien_2014,ASASSN-14li_10.1093/mnras/stv2486,Holoien_2016,Holoien_2019}, with the variance of the sample as the error, yielding $B/T=0.464\pm0.087$.  

We then use the total stellar mass derived using SED fitting (see section \ref{section:blast}) of  $\log_{10}\left(M_{\rm galaxy}/M_\odot\right) = 10.109^{+0.069}_{-0.078}$ and the $B/T$ ratios above to derive the bulge mass. 
We find a value of $\log_{10}(M_{\rm bulge}/M_\odot)= 9.994^{+0.071}_{-0.080}$ using the $g$-band measured $B/T$ and a value of $\log_{10}(M_{\rm bulge}/M_\odot)= 9.767^{+0.19}_{-0.19}$ using the average TDE host $B/T$.

Finally, adopting the black hole-bulge mass relation from \citet{BH_Bulge}, we infer SMBH masses of $\log_{10}(M_{\rm BH}/M_\odot)\approx 7.52^{+0.32}_{-0.32}$ and $\log_{10}(M_{\rm BH}/M_\odot)\approx 7.28 ^{+0.38}_{-0.38}$ for the $g$-band and averaged bulge-mass methods, respectively. All of our  SMBH mass estimates are summarized in Table~\ref{tbl:BH_mass_estimates}.

\begin{deluxetable*}{llll}
\tablecaption{SMBH mass estimates and corresponding Eddington ratios for the TDE second $g$-band peak and the AGN bolometric luminosities.}\label{tbl:BH_mass_estimates}
\tablehead{
\colhead{Method}  &
\colhead{$\log_{10}\left(\frac{M_{BH}}{M_\odot}\right)$} &
\colhead{$\log_{10}\left(\frac{L_{peak}}{L_{EDD}}\right)$}
&
\colhead{$\log_{10}\left(\frac{L_{AGN}}{L_{EDD}}\right)$}
}
\startdata
\texttt{MOSFiT\tablenotemark{a}} & $6.71^{+0.21}_{-0.21}$ & $-0.413_{-0.21}^{+0.21}$& $-1.37_{-0.35}^{+0.50}$ \\
\texttt{TDEmass} without cooling effect & $6.826^{+0.0097}_{-0.0016}$ & $-0.533_{-0.014}^{+0.013}$& $-1.56_{-0.21}^{+0.53}$ \\
\texttt{TDEmass} with cooling effect & $7.079^{+0.013}_{-0.016}$ & $-0.783^{+0.019}_{-0.018}$ &$-1.80_{-0.21}^{+0.53}$ \\
$M-\sigma_*$ relation\tablenotemark{b} & 6.18 $\pm$ 0.93 & $0.12_{-0.92}^{+0.92}$& $-0.78_{-0.99}^{+0.10}$ \\
Bulge-BH relation\tablenotemark{b,c} & $7.52^{+0.31}_{-0.32}$ & $-1.22^{+0.32}_{-0.31}$& $-2.15_{-0.44}^{+0.54}$\\
Bulge-BH relation\tablenotemark{b,d} & $7.28^{+0.38}_{-0.38}$ & $-0.99^{+0.38}_{-0.37}$& $-1.91_{-0.49}^{+0.57}$\\
\enddata
\tablenotetext{a}{Value includes systematic uncertainties from \citet{Mockler2019ApJ...872..151M}}
\tablenotetext{b}{\citet{Kormendy_2013}}
\tablenotetext{c}{$g$-band bulge ratio}
\tablenotetext{d}{TDE average bulge ratio}

\end{deluxetable*}

\section{Discussion}\label{section:discussion}
\subsection{Classification of AT\,2022csn as a TDE}
The spectroscopic properties of AT\,2022csn, such as the emergence of broad \ion{He}{2}, \ion{He}{1} and H$\alpha$ emission commonly observed in H+He TDEs \citep{van_Velzen_2020} strongly favor a TDE interpretation over that of a SLSN (Fig. \ref{fig:He_II_stacked}). Specifically, AT\,2022csn lacks the characteristic \ion{O}{2} absorption complexes of SLSNe-I \citep{Quimby_2018,Gal_Yam_2019} as well as the interaction-powered H-line profiles typical of SLSNe-II \citep{Gal_Yam_2012,Gal_Yam_2019_The_Most_Luminous_SNe}. 
Furthermore, the host-galaxy emission-line ratios place the host galaxy in the non-star-forming region of the BPT diagram (Fig. \ref{fig:BPT}), disfavoring the presence of massive young stellar populations required for producing SLSNe.  Its location in the $H\alpha$ EW- Lick $H\delta_A$ plane (Fig. \ref{fig:EW_Lick}) indicates a substantial Type-A star population implying a starburst within the past $\sim$Gyr, consistent with the post-starburst environments commonly associated with optical/UV TDEs \citep{French_2016}. 

\subsection{SMBH and Stellar Mass Estimates}

Using various methods to estimate the SMBH mass from both host-galaxy scaling relations and TDE emission models, we find that they all provide roughly consistent results around $\log_{10}(M_{BH}/M_\odot)\approx6.7$--$7.1$ (Table \ref{tbl:BH_mass_estimates}). This range is consistent with SMBH masses inferred for other optical/UV TDEs, but is in the higher end of the  distribution \citep{van_Velzen_2020}.

The inferred mass of the disrupted star differs substantially between the TDE models, with both \texttt{TDEMass} solutions (with and without cooling) favoring a significantly more massive star than \texttt{MOSFiT} ($M_\ast \approx 6.80^{+0.91}_{-0.74}$, $14.00^{+1.6}_{-1.2}$, and $1.00^{+0.03}_{-0.03}\ M_\odot$, respectively). It was recently shown that the stellar mass parameter is not well constrained by \texttt{MOSFiT} \citep{French2026}, and neither \texttt{MOSFiT} nor \texttt{TDEMass} were intended to produce double peaked light curves as seen here.

\subsection{Peculiar Photometric Properties of AT\,2022csn}

With a luminosity distance of $\sim$726 Mpc, AT\,2022csn stands out as one of the most distant and luminous optical/UV TDEs discovered so far\footnote{Though it is not the most luminous known optical-UV TDE \citep[e.g., AT\,2023vto;][]{Kumar_2024}.}, peaking at $L_{ peak}=2.487^{+0.073}_{-0.067}\times10^{44}\,{\rm erg\, s^{-1}}$. In addition, AT\,2022csn exhibits a pronounced double-peaked light curve, with the two $g$-band maxima separated by $18.30\pm2.84$ days. The double peak is seen across most photometric bands (Fig. \ref{fig:full_lightcurve}) and in the bolometric light curve (Fig. \ref{fig:lightcurve_fitting}). A third peculiar property of AT\,2022csn is that it lies at the low-temperature large-radius ends of the optical/UV TDE population (Fig. \ref{fig:lightcurve_fitting}).  

\subsection{AT\,2022csn as a TDE in an AGN}\label{section:TDE in AGN}

A possible explanation for the peculiar photometric behavior of AT\,2022csn is that it  occurred in a host galaxy harboring an AGN.
While we find no evidence of historic optical variability, the variability in the NEOWISE \citep{Mainzer_2011_NEOWISE} $W1$ and $W2$ photometry, the $W1-W2$ color of the host (Fig. \ref{fig:NEOWISE}) and the location of the host galaxy the AGN-Seyfert region of the BPT diagrams (Figure \ref{fig:BPT} and Figure \ref{fig:BPT_SII} in Appendix \ref{appendix:host_galaxy_spectral_fits}) all point to a likely AGN. 
The host spectrum of AT\,2022csn shows no evidence for emission lines expected from an AGN BLR \citep[FWHM $\gtrsim$ 2,000\,km\,s$^{-1}$;][]{Almeida_Ricci_2017NatAs...1..679R}, and instead exhibits only narrow emission lines with FWHM $\sim$100--150\,km\,s$^{-1}$ (Figure \ref{fig:PyQSOFit_DESI} in Appendix \ref{appendix:host_galaxy_spectral_fits}). We thus conclude that AT\,2022csn likely occurred in a Type II AGN. 

For a TDE in an AGN, interactions between the returning TDE stellar debris stream and a pre-existing AGN disk could perturb circularization and enhance shocks and/or outflows on larger characteristic scales than in non-AGN hosts. This may produce a more extended reprocessing layer, perhaps leading to larger effective blackbody radii and lower observed temperatures. The $\sim$18-day separation between the two lightcurve peaks corresponds to a light-travel distance of $\sim$0.02 pc, comparable to characteristic AGN  accretion disk scales \citep{accretion_disk2012MNRAS.422..129Z,Almeida_Ricci_2017NatAs...1..679R}, and therefore may be linked to disk-stream interaction. In the AGN unified model, Type II AGNs result from our viewing angle with respect to a toroidal obscuration that surrounds the central engine and the BLR \citep{unified_model1993ARA&A..31..473A}. In this case, it would be expected that the same obscuration would also surround the TDE. This raises the question of how it is possible to observe a TDE in a Type II AGN, since the TDE should be obscured by the same material obscuring the BLR.

Indeed, typical AGN torus sizes \citep[$\sim$0.1--10 pc;][]{Torus_size2008ApJ...681..141R,Netzer2015ARA&A..53..365N} are 2--4 orders of magnitude larger than the characteristic blackbody radii of TDEs (including AT\,2022csn). 
To estimate the torus scale in our system, we first derive the AGN bolometric luminosity from the narrow [\ion{O}{3}] $\lambda5007$ emission line measured with \texttt{PyQSOFit} \citep{pyqsofit_1_2018ascl.soft09008G,pyqsofit_2_2019ApJS..241...34S}, correcting for extinction using the measured H$\alpha$/H$\beta$ ratio following \citet{Bassani_1999}, and applying the bolometric conversion of \citet{Lamastra_2009}, we find $L_{AGN}=2.35^{+2.90}_{-1.08}\times10^{43}\,{\rm erg\,s^{-1}}$, with uncertainties estimated through a combination of Monte Carlo sampling and error propagation (see Appendix \ref{Appendix:Monte Carlo Sampling} for more details). In order to isolate the torus emission, we model the host-galaxy SED ourselves (Appendix \ref{appendix:Prospector SED}) using \texttt{Prospector} with the \texttt{Prospector-$\alpha$} framework \citep{Leja_2017,Johnson_2021}.
From this analysis, we infer a torus scale height of $0.59^{+0.78}_{-0.33}\ \rm pc$ (uncertainties are derived via Monte Carlo propagation; see Appendix \ref{Appendix:Monte Carlo Sampling} and Appendix \ref{appendix:Prospector SED}), which lies well within the known AGN torus size range and several orders of magnitude larger than the TDE photosphere. To test whether the observed $W1$ and $W2$ photometry and color could be explained by stellar emission alone, we re-ran the \texttt{Prospector-$\alpha$} model with the AGN component disabled and included the IRAS upper limits in order to constrain the model (Appendix \ref{appendix:Prospector SED}). The resulting fit (Figure \ref{fig:prospector_no_agn} in Appendix \ref{appendix:Prospector SED}) shows that the NEOWISE measurements cannot be reproduced without a torus component. Therefore, we conclude that an explanation without a torus contribution is disfavored.
Below, we discuss possible explanations for how a TDE might remain observable in a host galaxy displaying Type II AGN features.

\subsubsection{An Outshined Type I AGN}\label{section:outshined_type_I}
One possible explanation is that instead of a Type II AGN, the host is viewed as a Type I AGN whose accretion disk and BLR are outshined in the optical/UV by the stellar emission of the galaxy. We estimate the expected AGN continuum emission at 5100\,\AA{} ($L_{5100}$) from our inferred $L_{ AGN}$ following \citet{Richards_2006} once using the bolometric correction from \citet{Netzer_2019} and once using the bolometric correction from \citet{Wu_2022}. For each $L_{5100}$ estimate, we construct a standard accretion-disk spectrum with a $-7/3$ power-law slope and estimate the corresponding broad H$\alpha$ and H$\beta$ luminosities using the relations of \citet{2005ApJBHHalpha}. For the bolometric corrections of \citet{Netzer_2019} and \citet{Wu_2022}, respectively, we obtain  $L_{H\alpha}=1.18^{+2.94}_{-0.75}\times10^{40}\, {\rm erg\,s^{-1}}$, $L_{H\beta}=3.64^{+8.78}_{-2.28}\times10^{39}\, {\rm erg\,s^{-1}}$ and $L_{H\alpha}=7.48^{+13.65}_{-4.28}\times10^{40}\, {\rm erg\,s^{-1}}$, $L_{H\beta}= 2.219^{+3.94}_{-1.26}\times10^{40}\, {\rm erg\,s^{-1}}$. In addition, we directly convert our measured [\ion{O}{3}] luminosity, $L_{[O_{III}]}=7.43\pm1.24\times10^{40}\ {\rm erg\,s^{-1}}$, to $L_{H\alpha}=3.44^{+7.67}_{-2.38}\times10^{41}\ {\rm erg\,s^{-1}}$ using the relations of \citet{Stern_Laor_2012}. Uncertainties were estimated through a combination of Monte Carlo sampling and error propagation (see Appendix \ref{Appendix:Monte Carlo Sampling}). We plot these components together with our \texttt{Prospector-$\alpha$} SED fit in Figure \ref{fig:Prospector} in Appendix \ref{appendix:Prospector SED}. For both $L_{5100}$ estimates, the expected accretion-disk and BLR contribution are  subdominant to the stellar continuum across nearly all bands, except for the GALEX $FUV$ band in the case of the \citet{Wu_2022} bolometric correction. The main exception is the \citet{Stern_Laor_2012} broad H$\alpha$ estimate, which would be detectable above the stellar continuum. For both bolometric corrections, the mid-infrared $W1$ and $W2$ bands require an additional torus component.

In order to quantify this comparison, we estimate an upper limit on the broad H$\alpha$ emission from the pre-flare DESI host spectrum and compare it with the expected broad H$\alpha$ luminosities described above. We estimate the upper limit by first measuring the typical $1\sigma$ noise level around H$\alpha$ from the median absolute deviation of the continuum subtracted DESI spectrum in two adjacent emission-line-free continuum windows, 6150--6250\,\AA{} and 6950--7150\,\AA{}. We then take the upper limit on the broad H$\alpha$ flux to be the flux of a Gaussian with peak amplitude equal to this noise level and a FWHM of 5,000\,${\rm km\,s^{-1}}$. We adopt a FWHM of 5,000 ${\rm km\,s^{-1}}$ for the broad H$\alpha$ upper limit as an intermediate value within the range implied by our expected BLR line widths. Specifically, the \citet{Stern_Laor_2012} [\ion{O}{3}]-based estimate gives FWHM $\sim3,000\ {\rm km\,s^{-1}}$, while the estimates based on $L_{5100}$ using the bolometric corrections of \citet{Wu_2022} and \citet{Netzer_2019} give FWHM $\sim6,000$ and $\sim10,000\ {\rm km\,s^{-1}}$, respectively. The adopted value is therefore representative of the expected range and is also consistent with typical BLR line widths. Converting the resulting $3\sigma$ flux limit to luminosity and applying the same Balmer-decrement extinction correction used for  the [\ion{O}{3}] luminosity gives $L_{\rm H\alpha,broad}< 2.1\times10^{41}\ {\rm erg\,s^{-1}}$. The two $L_{5100}$-based broad H$\alpha$ estimates are below to this limit, indicating that such a component could remain hidden in the spectral noise or be outshined by the host galaxy continuum. In contrast, the \citet{Stern_Laor_2012} prediction implies a marginally detectable broad H$\alpha$. All estimated broad H$\alpha$ components are illustrated on top of the observed spectrum in Figure \ref{fig:Prospector} in Appendix \ref{appendix:Prospector SED}. 

In addition, we evaluate the size of the BLR using our estimated luminosities of H$\alpha$ and the relations of \citet{Cho_2023}. For the luminosity estimates of \citet{Stern_Laor_2012}, \citet{Netzer_2019}, and \citet{Wu_2022}, we obtain $R_{\rm BLR}=6.33^{+10.37}_{-3.94}\times10^{-3}$, $8.09^{+14.37}_{-4.85}\times10^{-4}$, and $2.49^{+3.75}_{-1.42}\times10^{-3}\ {\rm pc}$, respectively. Uncertainties were estimated through a combination of Monte Carlo sampling and error propagation (see Appendix \ref{Appendix:Monte Carlo Sampling}). These radii are consistent with expectations for a weak or low luminosity broad line AGN.

We also estimate the hard X-ray luminosity expected from the narrow [\ion{O}{3}] emission and compare it with the pre-flare non-detections from the extended ROentgen Survey with an Imaging Telescope Array \citep[eROSITA;][]{erosita2021A&A...647A...1P}. The eROSITA non-detections imply a 3$\sigma$ upper limit of $L_{2-10,{\rm keV}}\lesssim5.1\times10^{42}\ {\rm erg\,s^{-1}}$ (see Appendix \ref{appendix:X-ray upper limits}). Using our measured [\ion{O}{3}] luminosity, and the relation of \citet{Ueda_2015}, we infer an expected $L_{2-10\,{\rm keV}}\sim1.3\times10^{43}\ {\rm erg\,s^{-1}}$ (Appendix \ref{appendix:X-ray upper limits}). The inferred luminosity is therefore above the eROSITA limit by $\sim0.4$ dex, making the comparison marginal, although this difference is smaller than the observed scatter of local hard X-ray selected AGNs reported by \citet{Ueda_2015}. 
Such a scenario can naturally reconcile the observed TDE with the outshined Type I AGN interpretation, as the system is not viewed through an obscuring torus.

\subsubsection{True Type II AGN}\label{section:true_type_II}
Another possible scenario to explain how a TDE is visible in a Type II AGN is with a True Type II AGN. True Type II AGNs show no broad lines but also no signs of obscuration (such as little or no X-ray absorption), suggesting that the absence of broad lines is intrinsic rather than due to line of sight obscuration \citep[e.g.][]{Bianchi_2008,Brightman_2008,Bianchi_2012,Netzer2015ARA&A..53..365N,Elitzur_2016}. Although the interpretation of True Type II AGNs remains debated, as some candidate systems may still be consistent with weak broad H$\alpha$ emission that falls below the detection limits of the available spectra \citep{Stern_Laor_2012_a,Stern_Laor_2012}, such systems have been suggested to constitute $\sim$30\% of the Type II AGN population in the local Universe \citep{Panessa_2002,Marinucci_2012,Merloni_2014}. Many of them are characterized by low bolometric luminosities and low Eddington ratios of order $\sim$1\%, consistent with a weak or absent accretion disk \citep{Tran2001ApJ...554L..19T,Elitzur_HO_2009ApJ...701L..91E,Elitzur_2016}, although examples with higher luminosities and Eddington ratios have also been reported and discussed theoretically \citep[e.g.,][]{Ho_2012,Miniutti_2013,Elitzur_2016}. 

Our inferred value of $L_{AGN}$ and Eddington ratio (Table \ref{tbl:BH_mass_estimates}) lie within (but toward the upper end of) the range typically associated with True Type II AGNs. At face value, a True Type II interpretation may seem to disfavor the presence of a torus, which is expected to disappear together with the BLR. In that case, the inability of a torus free model to reproduce the observed $W1$ and $W2$ photometry (see Appendix \ref{appendix:Prospector SED}) would appear to be in tension with this scenario. However, \citet{Netzer2015ARA&A..53..365N} argue that many Type II AGNs lack sufficient high-density gas to produce a detectable BLR while still retaining a central torus, which could reconcile the True Type II picture with our inferred mid-infrared emission.

One proposed explanation for True Type II AGNs is a rapid AGN fading (or ``shutdown"), in which the BLR and disk emission fade on short timescales while narrow-line emission persists because of light-travel and recombination time \citep{Binette_1987,Schawinski_2010}.
To explore this possibility for our AGN, we estimate the distance between the NLR and the central engine for the host of AT\,2022csn using the relation of \citet{Bennert_2002} and the extinction-corrected luminosity of [\ion{O}{3}], with uncertainties derived from Monte Carlo sampling (Appendix \ref{Appendix:Monte Carlo Sampling}). We obtained $$R_{ NLR}= 2,311^{+2,436}_{-1,231}\ {\rm ly}$$ This is the time window in years during which the narrow line region would remain visible after the AGN significantly weakened. Given recent suggestions for enhanced TDE rates in AGN-host galaxies \citep{Stone_2018,Kaur_2025ApJ...979..172K}, reaching $\sim10^{-4} - 10^{-3}\ {\rm events\ galaxy^{-1}\ yr^{-1}}$, it is not unlikely for a TDE to occur in this time window.

\subsubsection{Ionization From a Previous TDE}
Some post-starburst galaxies exhibit extended emission-line regions consistent with ionization from past AGN activity, which can be interpreted as evidence for fading AGN \citep{French_2023} or for previous TDEs \citep{Mummery_2025}. \citet{Mummery_2025} have shown that long-lived TDE disks can produce sufficient ionizing flux to power extended emission-line regions out to $r \sim 10^{4}$\,ly. In this scenario, a TDE occurring a few thousand years before AT\,2022csn could have photoionized the NLR instead of an AGN. If the host is indeed a post-starburst galaxy, as suggested by its strong $H\delta_A$ absorption (Figure \ref{fig:EW_Lick}), then the enhanced TDE rates observed in such galaxies \citep[of order $\sim 10^{-2}$--$10^{-3}\ {\rm events\ galaxy^{-1}\ yr^{-1}}$; e.g.][]{Decker_arcavi2017ApJ...835..176F,French_2020} would allow for such a scenario.
Indeed, extended emission-line regions have been identified in some post-starburst galaxies that hosted TDEs \citep{Wevers_2024,Pursiainen_2026}, potentially tracing ionization from previous tidal disruptions. For the typical TDE disk light echo case presented by \citet{Mummery_2025}, the predicted luminosities are roughly an order of magnitude below those we observe. However, our measured values can still be reproduced using alternative disk and environmental parameters within the range explored by \citet{Mummery_2025}.

\subsubsection{Shocked Post Starburst Galaxy}
We have shown that the host galaxy of AT\,2022csn lies also within the region occupied by SPOGs in both the BPT diagrams (Figure \ref{fig:BPT} and Figure \ref{fig:BPT_SII} in Appendix \ref{appendix:host_galaxy_spectral_fits}) and in the H$\alpha$ EW vs. Lick H$\delta_A$ absorption plane (Fig. \ref{fig:EW_Lick}). SPOGs are thought to represent galaxies transitioning from star-forming blue to quiescent red, likely at an earlier evolutionary stage than classical E+A/K+A post-starburst galaxies \citep{French_2018}. They are characterized by strong Balmer absorption from A-type stars, indicating significant star formation within the past $\sim$Gyr, but unlike traditional E+A or K+A selections, the SPOG classification allows for the presence of H$\alpha$ emission, which is attributed to shocks or AGN activity rather than star formation \citep{Altalo_2016ApJS..224...38A}. Several studies have suggested enhanced TDE rates in post-starburst galaxies \citep[e.g.][]{Arcavi_2014,Decker_arcavi2017ApJ...835..176F,French_2020}, which could imply intrinsically elevated TDE rates in SPOGs as well, even if the observed rates may be suppressed by additional dust obscuration \citep{French_2018}. Under this interpretation, our TDE is visible because the host galaxy is a SPOG, with the H$\alpha$ emission coming from shocks rather than from an AGN. We note, however, that \citet{Altalo_2016ApJS..224...38A} caution against interpreting the relevant BPT regions as arising solely from shocks, and emphasize that some contribution to the H$\alpha$ emission may still come from an AGN.

We compare our observed H$\beta$ to [\ion{O}{3}] $\lambda5007$ line ratios to those from the models of \citet{Allen_2008}. We find that the observed line ratios are not consistent with a pure shock component \citep[Table 6 in][]{Allen_2008}, but are consistent with both a precursor component with shock velocities of $v_s\sim$ 200--300\,$\rm km\,s^{-1}$ \citep[Table 7 in][]{Allen_2008} and a shock plus precursor component with $v_s \sim$ 300--400\,$\rm km\,s^{-1}$ \citep[Table 8 in][]{Allen_2008}. The precursor origin for the lines would require an emitting radius of $\sim$ 3.7--6.4\,kpc, while the shock plus precursor origin requires a smaller emitting radius of $\sim$ 1.7--2.5\,kpc. Both inferred radii are consistent with the host-galaxy Petrosian radius of 6.6 kpc derived from SDSS DR13 \citep{Albareti_2017}.

\subsubsection{Offset TDE in a SMBH Binary}
Another possible scenario is that AT\,2022csn is a TDE around an SMBH which is part of a binary, with the companion being the active SMBH. In that case the TDE could be sufficiently offset from the AGN obscuration, and perhaps observed as offset from its host center. Only a small number of TDEs offset from their host center have been identified to date, including 3XMMJ2150 \citep[$\sim$12.5 kpc off-nuclear;][]{Offset_TDE_3XMM2018NatAs...2..656L}, AT\,2024tvd \citep[$\sim$0.8 kpc off-nuclear;][]{AT2024tvd2025ApJ...985L..48Y}, and TDE 2025abcr \citep[10.3 kpc off-nucler][]{Stein_2026}.
Using the ZTF discovery positions of AT\,2022csn and the centroid of its host galaxy SDSS J072854.92+265349.3 \citep{fremling2022TNSTR.452....1F}, we estimate a projected offset of $0.38\pm0.25$ kpc. While this will place the TDE outside the obscuring torus of a nuclear Type II AGN, the offset measurement is not statistically significant.

Regardless of the offset, enhanced TDE rates are expected in SMBH binaries systems, as dynamical mechanisms such as eccentric Kozai-Lidov \citep{Kozai1962AJ.....67..591K,Lidov1962P&SS....9..719L} oscillations and chaotic three-body interactions can efficiently drive stars to loss-cone orbits, producing a temporary increase in the disruption rate \citep{Chen_2009,Chen_2011,Mockler_2023}. In addition, perturbations from a binary companion can introduce substantial fluctuations and gaps in the stellar mass fallback rate, potentially imprinting irregular structure on the optical/UV light curve \citep{Wen_2024}, perhaps explaining the double peak seen in AT\,2022csn.

\subsubsection{TDE on a Nearly Circular Orbit}
A further scneario to consider is Roche lobe overflow from a star on a nearly circular orbit around the SMBH, as proposed by \citet{Linial_2024}. In this scenario, X-ray emission from sustained mass transfer onto the black hole may precede the main optical flare on time scales of years to a decade. Such emission might provide sufficient ionizing radiation to power the observed NLR, though may be undetected given the large distance to the source. The relatively short delay between the ionizing X-rays and the main optical TDE flare could perhaps produce the observed narrow lines if the NLR lies close to our line of sight to AT\,2022csn.

\subsection{Other TDEs in Narrow Emission Lines Hosts}\label{section:other_TDEs_in_AGNs}
We searched for other robust TDEs in the \citet{van_Velzen_2020} sample that are associated with narrow emission line host galaxies, and identified AT\,2019ahk \citep{Holoien_2019}, see also \citet{Merloni_2015} for a possible TDE flare in an AGN exhibiting a significantly longer lived flare and  different spectral characteristics compared to an AT\,2022csn-like event. AT\,2019ahk is a TDE showing both H and \ion{He}{1} emission but no detectable \ion{He}{2}. A 6\,dF \citep{6DF_2009MNRAS.399..683J} archival spectrum of its host obtained from NED exhibits narrow emission lines including [\ion{O}{3}] $\lambda 5007$, H$\alpha$, H$\beta$, and [\ion{N}{2}] $\lambda6583$. AT\,2019ahk show similarly low temperatures as those of AT\,2022csn. While the blackbody radius of AT\,2019ahk appears slightly smaller than that of AT\,2022csn, it remains large compared to the rest of the TDE sample \citep{van_Velzen_2021}. A key observational difference is that AT\,2019ahk displays a smooth, single-peaked light curve, whereas AT\,2022csn exhibits a pronounced double peak.

The host galaxies of the two events are very similar. In both the BPT diagnostics and the H$\alpha$ EW-H$\delta_A$ Lick absorption phase space (Figures \ref{fig:BPT} and \ref{fig:EW_Lick}), the host of AT\,2019ahk lies near that of AT\,2022csn. The host-galaxy SED-fitting parameters derived from \texttt{Blast} are also similar for both galaxies (Fig. \ref{fig:Blast_hist}). 

In addition, \citet{Holoien_2019} reported a pre-TDE $3\sigma$ absorbed X-ray upper limit of $L_{0.3-10{\rm keV}}\lesssim2.7\times10^{42}\,{\rm erg\, s^{-1}}$ based on the ROSAT All-Sky Survey \citep{ROSAT1999A&A...349..389V}, that is also consistent with the weak AGN expected in True Type II systems \citep{Tran2001ApJ...554L..19T,Elitzur_HO_2009ApJ...701L..91E,Elitzur_2016,Ricci_2017}. 

Taken together, AT\,2022csn and AT\,2019ahk represent a subset of TDEs occurring in narrow emission line galaxies consistent with Type II AGNs, characterized by relatively low temperatures and unusually large blackbody radii compared to the bulk of the TDE sample \citep{van_Velzen_2021}. This raises the possibility that either low temperatures and large radii are a result of the interaction with an AGN disk, or that such properties make TDEs in AGN easier to detect.

\subsection{Other Double Peaked TDEs}
Only a small number of TDEs with double-peaked light curves have been reported (namely, AT\,2019ehz; \citealt{van_Velzen_2021}, AT\,2020acka; \citealt{2020ackaTNSCR.262....1H}, AT\,2021uqv; \citealt{2021uqvTNSCR3411....1Y}, and AT\,2019baf; \citealt{Yao_2023}).
AT\,2020acka was also noted as unusually luminous compared to the broader TDE population \citep{yao2025opticallyoverluminoustidaldisruption}. However, in these events the separation between peaks is typically of order $\sim$200--300 days which is much longer than in AT\,2022csn, which has a peak separation of only $\sim$18 days. This suggests that the double-peaked behavior in AT\,2022csn may arise from a different mechanism than those proposed for the longer-timescale cases, which are often discussed in the context of repeating partial disruptions \citep[e.g.][]{Zhong_2025}.

\section{Summary and Conclusions}\label{section:summary}
AT\,2022csn is among the most luminous and distant optical/UV TDEs observed to date with a peak bolometric luminosity of $L_{ peak}=2.487^{+0.073}_{-0.067}\times10^{44}\,{\rm erg\,s^{-1}}$ and a luminosity distance of $\sim$726 Mpc. The spectroscopic properties of AT\,2022csn, such as broad \ion{He}{2}, \ion{He}{1} and H$\alpha$ emission (Fig. \ref{fig:He_II_stacked}) are commonly observed in H+He TDEs \citep{van_Velzen_2020}. The host-galaxy spectrum shows strong H$\delta$ absorption consistent with a significant starburst within the last $\sim$Gyr (Fig. \ref{fig:EW_Lick}), which is also commonly observed in other optical/UV TDEs. However, here we find also narrow emission lines that place the host galaxy in the AGN-Seyfert region of the BPT diagrams (Figure \ref{fig:BPT} and Figure \ref{fig:BPT_SII} in Appendix \ref{appendix:host_galaxy_spectral_fits}). 

AT\,2022csn has several unusual photometric properties: it shows a pronounced double-peaked light curve in most bands and in its bolometric light curve, with peaks separated by 18.30 $\pm$ 2.84 days (in the $g$-band), and it shows low temperatures and high radii relative to other H/He TDEs. These peculiar properties might be related to the fact that the host galaxy of AT\,2022csn harbors an AGN. Indeed, the $\sim$18 day peak separation corresponds to a light-travel scale of $\sim$0.02 pc, which is comparable to AGN disk scales \citep{accretion_disk2012MNRAS.422..129Z,Almeida_Ricci_2017NatAs...1..679R}, potentially linking the light-curve structure to interaction between the TDE debris stream and the pre-existing AGN disk.

The lack of broad lines in the host spectrum (Figure \ref{fig:Stacked_spectra_norm} and Figure \ref{fig:PyQSOFit_DESI} in Appendix \ref{appendix:host_galaxy_spectral_fits}) imply a Type II AGN. However, within the unified AGN model, such a system would typically obscure the region around the SMBH \citep{unified_model1993ARA&A..31..473A}, from which the TDE emission originates, as the TDE photosphere ($\sim5\times10^{-4}$ pc) lies well within the inferred BLR ($\sim10^{-3}$ pc) and torus ($\sim0.6$ pc), while only the NLR extends to much larger scales ($\sim700$ pc). We consider a few explanations as to how a TDE might still be visible in such an AGN. One possibility is that the host is an outshined Type I AGN, in which standard accretion-disk, BLR, and torus components are present but the optical/UV AGN signatures are outshined by the host-galaxy stellar emission. Another possible explanation, is a True Type II AGN, where the absence of broad lines is intrinsic rather than due to obscuration \citep{Bianchi_2008,Elitzur_2016}. The AGN luminosity and X-ray non-detections are consistent with this scenario. The observed narrow emission may not originate from an AGN at all, but instead be an ionization echo from a previous TDE that photoionized the surrounding gas. Another option is that the host galaxy is a SPOG, i.e. that the narrow lines originate from post-starburst shocks. Alternatively, AT\,2022csn might be a TDE around a SMBH that is a member of a binary, with the companion SMBH hosting the AGN, and the two SMBHs separated enough that the TDE is not obscured. Finally, Roche lobe overflow from a star on a nearly circular orbit around an SMBH could generate sustained X-ray emission prior to the main flare, potentially ionizing the NLR on the relevant timescale given a favorable geometry.

We identified AT\,2019ahk \citep{Holoien_2019} as a close analog in both host-galaxy and blackbody properties, suggesting a possible subset of low-temperature, high-radius TDEs in narrow emission line galaxies. Together, these events motivate future searches for additional TDEs in AGN-like hosts, and highlight the potential of TDEs in AGN to constrain both stellar disruption emission mechanisms and the structure of weak or transitional AGN.

We thank A. Mummery., I. Linial and J. Stern for insightful discussions, A. Boestroem, J. Farah and L. Makrygianni for help using the \texttt{lcogtsnpipe} pipeline.
Y.D. acknowledges support from the United States - Israel Binational Science Foundation (BSF; grant number 2024812).
I.A. acknowledges support from the European Research Council (ERC) under the European Union’s Horizon 2020 research and innovation program (grant agreement number 852097), from the Israel Science Foundation (grant number 2752/19), from the BSF (grant number 2024812), and from the Pazy foundation (grant number 216312).
The authors thank the Yukawa Institute for Theoretical Physics at Kyoto University. Discussions during the YITP long-term workshop YITP-T-26-02 on ``Multi-Messenger Astrophysics in the Dynamic Universe'' were useful to this work.
\texttt{IRAF} is written and supported by the National Optical Astronomy Observatories, operated by the Association of Universities for Research in Astronomy, Inc. under cooperative agreement with the National Science Foundation.
The ZTF forced-photometry service was funded under the Heising-Simons Foundation grant
$\#12540303$ (PI: Graham).
These results made use of the Lowell Discovery Telescope (LDT) at Lowell Observatory. Lowell is a private, non-profit institution dedicated to astrophysical research and public appreciation of astronomy and operates the LDT in partnership with Boston University, the University of Maryland, the University of Toledo, Northern Arizona University and Yale University.
The upgrade of the DeVeny optical spectrograph has been funded by a generous grant from John and Ginger Giovale and by a grant from the Mt. Cuba Astronomical Foundation.
NBM and GGM acknowledge funding from the European Union (ERC, CET-3PO, 101042610), financial support from grant CEX2024-001451-M funded by MICIU/AEI/10.13039/501100011033, and support through research project PID2024-155585NA-I00 funded by MICIU/AEI /10.13039/501100011033.
BT acknowledges support from the ERC under the European Union’s Horizon 2020 research and
innovation program (grant agreement No. 950533)

Views and opinions expressed are, however, those of the author(s) only and do not necessarily reflect those of the European Union or the European Research Council Executive Agency. Neither the European Union nor the granting authority can be held responsible for them.

This work is supported by the National Science Foundation under Cooperative Agreement PHY-2019786 (The NSF AI Institute for Artificial Intelligence and Fundamental Interactions \footnote{\url{http://iaifi.org/}})
This research uses services or data provided by the Astro Data Lab, which is part of the Community Science and Data Center (CSDC) Program of NSF NOIRLab. NOIRLab is operated by the Association of Universities for Research in Astronomy (AURA), Inc. under a cooperative agreement with the U.S. National Science Foundation.
This research uses services or data provided by the SPectra Analysis and Retrievable Catalog Lab (SPARCL) and the Astro Data Lab, which are both part of the Community Science and Data Center (CSDC) Program of NSF NOIRLab. NOIRLab is operated by the Association of Universities for Research in Astronomy (AURA), Inc. under a cooperative agreement with the U.S. National Science Foundation.
This research used data obtained with the Dark Energy Spectroscopic Instrument (DESI). DESI construction and operations is managed by the Lawrence Berkeley National Laboratory. This material is based upon work supported by the U.S. Department of Energy, Office of Science, Office of High-Energy Physics, under Contract No. DE-AC02-05CH11231, and by the National Energy Research Scientific Computing Center, a DOE Office of Science User Facility under the same contract. Additional support for DESI was provided by the U.S. National Science Foundation (NSF), Division of Astronomical Sciences under Contract No. AST-0950945 to the NSF’s National Optical-Infrared Astronomy Research Laboratory; the Science and Technology Facilities Council of the United Kingdom; the Gordon and Betty Moore Foundation; the Heising-Simons Foundation; the French Alternative Energies and Atomic Energy Commission (CEA); the National Council of Humanities, Science and Technology of Mexico (CONAHCYT); the Ministry of Science and Innovation of Spain (MICINN), and by the DESI Member Institutions: www.desi.lbl.gov/collaborating-institutions. The DESI collaboration is honored to be permitted to conduct scientific research on I’oligam Du’ag (Kitt Peak), a mountain with particular significance to the Tohono O’odham Nation. Any opinions, findings, and conclusions or recommendations expressed in this material are those of the author(s) and do not necessarily reflect the views of the U.S. National Science Foundation, the U.S. Department of Energy, or any of the listed funding agencies. D.J. acknowledges support from the Spanish Agencia Estatal de Investigaci\'on del Ministerio de Ciencia, Innovaci\'on y Universidades (MCIU/AEI) under grants PID2022-136653NA-I00 and CNS2023-143910  (DOI:10.13039/501100011033). Based on observations made with the Nordic Optical Telescope, owned in collaboration by the University of Turku and Aarhus University, and operated jointly by Aarhus University, the University of Turku and the University of Oslo, representing Denmark, Finland and Norway, the University of Iceland and Stockholm University at the Observatorio del Roque de los Muchachos, La Palma, Spain, of the Instituto de Astrof\'isica de Canarias. The NOT data presented here were obtained with ALFOSC, which is provided by the Instituto de Astrof\'isica de Andalucia (IAA) under a joint agreement with the University of Copenhagen and NOT.
This research has made use of the NASA/IPAC Extragalactic Database, which is funded by the National Aeronautics and Space Administration and operated by the California Institute of Technology.
This research has made use of data and software provided by the High Energy Astrophysics Science Archive Research Center (HEASARC), which is a service of the Astrophysics Science Division at NASA/GSFC.
Based on observations collected at the European Southern Observatory under ESO program 108.2262.001

\facilities{HST(STIS), Swift(XRT and UVOT), AAVSO, CTIO:1.3m, CTIO:1.5m, CXO,LDT(DeVeny), NOT(ALFOSC), IRSA, IRAS, WISE}

\software{astropy \citep{2013A&A...558A..33A,2018AJ....156..123A,2022ApJ...935..167A},  
          Cloudy \citep{2013RMxAA..49..137F}, 
          Source Extractor \citep{1996A&AS..117..393B}, SPARCL \citep{SPARCL_juneau2025sparclspectraanalysisretrievable}
          }

\appendix
\section{\texttt{MOSFiT} Posterior Distributions}\label{appendix:MOsfit corner}
\begin{figure*}[ht!]
\includegraphics[width=\textwidth]{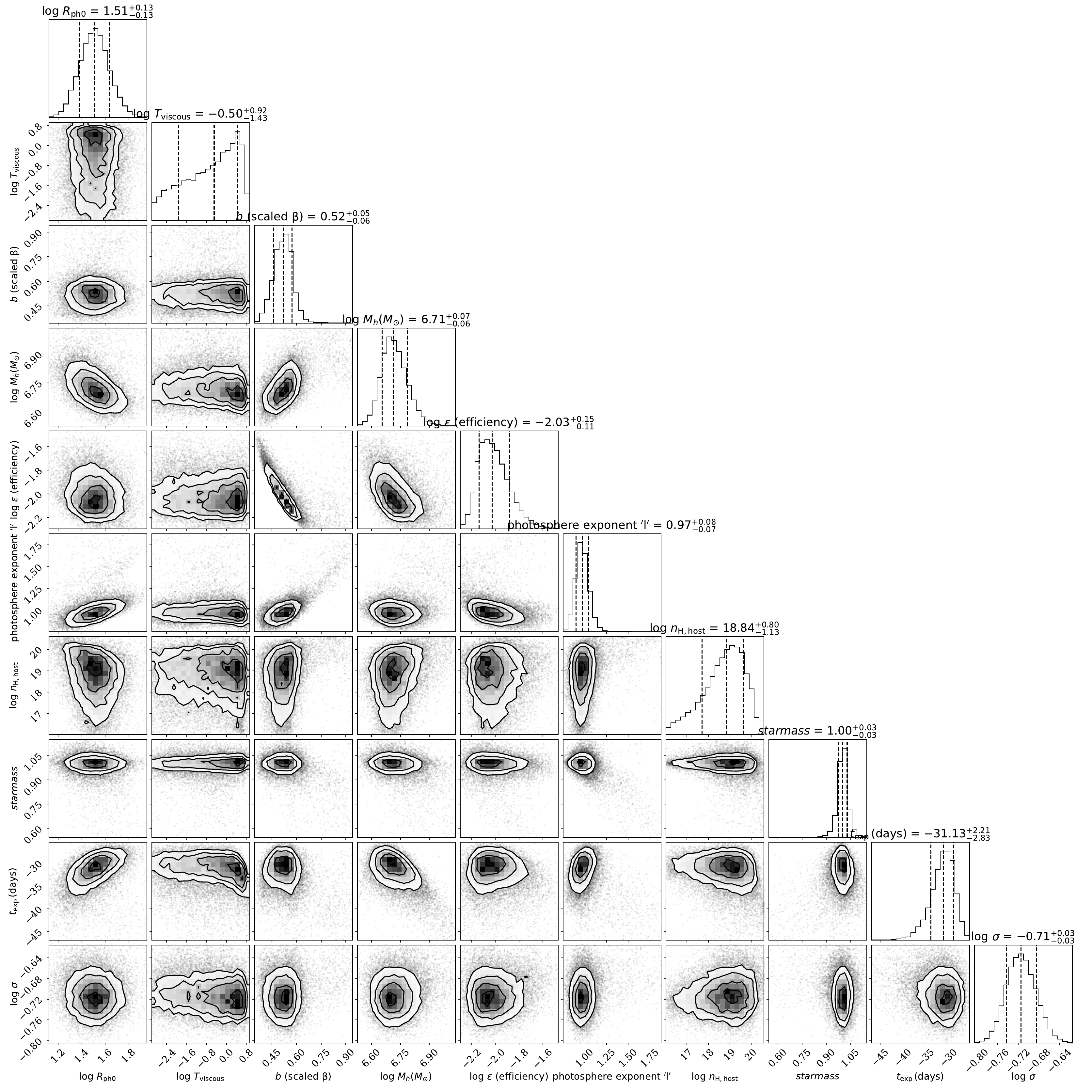}
\caption{Corner plot showing the posterior parameter distributions for the MOSFiT model fit. The 16th, 50th and 84th percentiles are marked with dashed vertical lines.}
\label{fig:mosfit_corner}
\end{figure*}

The posterior distributions and associated corner plot from the reprocessing emission model fit by \texttt{MOSFiT} are shown in Figure \ref{fig:mosfit_corner}.

\section{Host-Subtracted TDE Spectra}\label{appendix:Stacked_spectra_host}
\begin{figure}[ht!]
\centering
\includegraphics[width=0.5\textwidth]{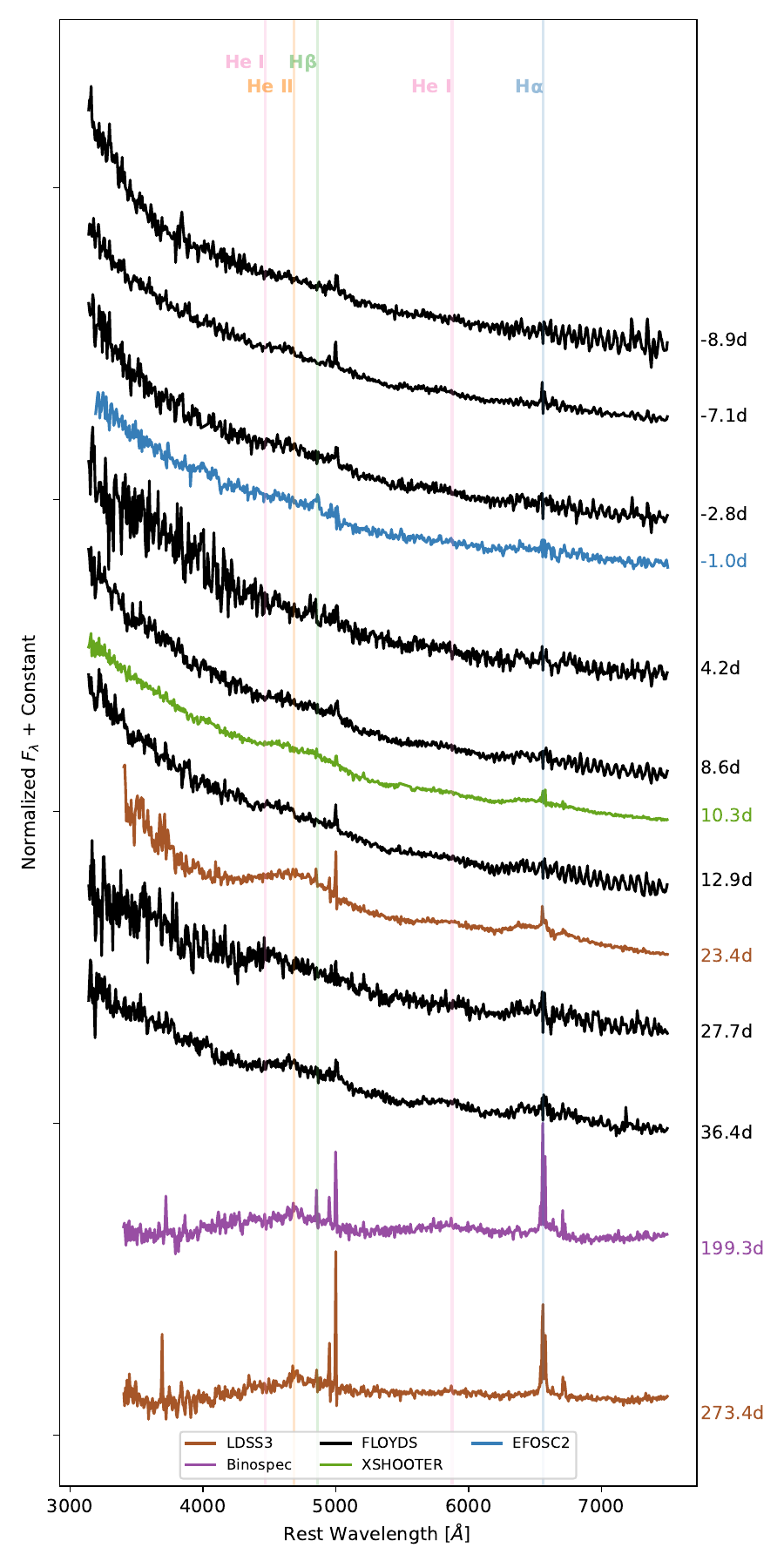}
\caption{Photometry-calibrated and host-subtracted spectra of AT\,2022csn. Rest-frame days with respect to the first $g$-band peak at 59648.17 MJD are noted. Remaining narrow emission lines are due to imperfect host subtraction.}
\label{fig:Stacked_spectra_host}
\end{figure}

The host-subtracted spectra of AT\,2022csn, described in Section \ref{section:analysis_spectroscopy}, are presented in Figure \ref{fig:Stacked_spectra_host}.

\section{Archival light curve}\label{appendix:ZTF ATLAS archival}
\begin{figure*}[ht!]
\includegraphics[width=\textwidth]{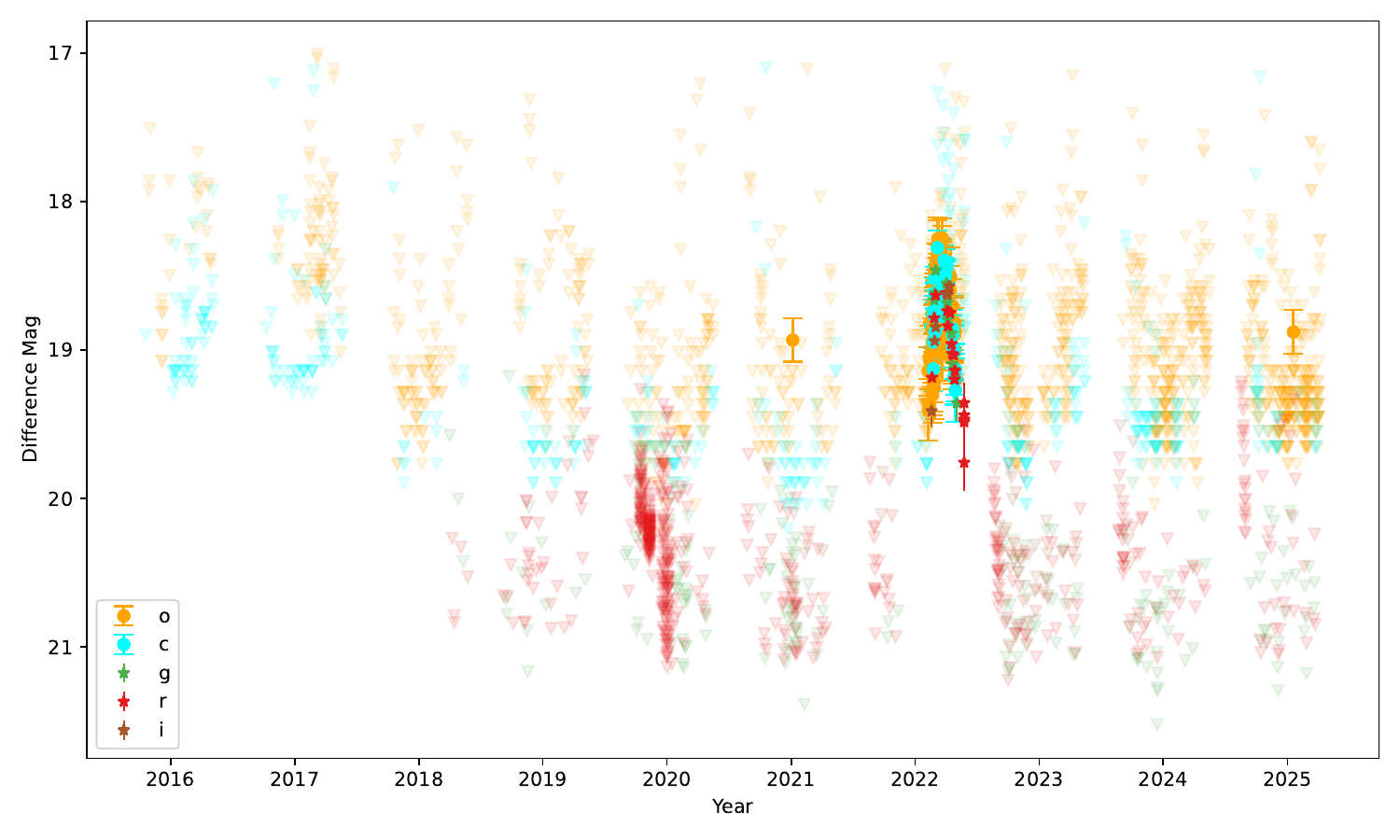}
\caption{Forced ATLAS and ZTF photometry at the position of AT\,2022csn. Triangles denote $5\sigma$ non-detection upper limits.}
\label{fig:archival_lightcurve}
\end{figure*}

The ATLAS and ZTF forced photometry at the position of AT\,2022csn are shown in Figure \ref{fig:archival_lightcurve}. We find no significant emission at the $5\sigma$ level either prior to or following the TDE.

\section{Monte Carlo Sampling Process}\label{Appendix:Monte Carlo Sampling}
In each realization, the narrow [\ion{O}{3}], H$\alpha$, and H$\beta$ luminosities were drawn from normal distributions centered on the best-fit values obtained with \texttt{PyQSOFit}, with standard deviations given by their measured uncertainties. 
Derived quantities that depend directly on the extinction-corrected [\ion{O}{3}] luminosity, such as the torus height and the NLR distance, were calculated for each realization. For quantities derived from $L_{AGN}$ through literature scaling relations (e.g., AGN continuum luminosity at 5100 \AA{}, expected broad-line luminosities, and $R_{BLR}$), uncertainties were obtained by propagating both the Monte Carlo uncertainty on $L_{AGN}$ and the quoted uncertainties of the corresponding relations.
\section{Fits to the Host Galaxy Photometry}\label{appendix:host_galaxy_photometry}
\subsection{\texttt{Blast} SED fitting parameters}\label{appendix:Blast_parameters}
The \texttt{Blast} SED fitting parameters are:
\begin{enumerate}
\item The logarithm of the stellar mass, age and metalicity of the galaxy - $\log_{10}(M_\ast/M_\odot)$, $\log_{10}\left(\frac{\mathrm{Age}_\ast}{\mathrm{yr}}\right)$ and $\log_{10}\left(\frac{Z_\ast}{Z_\odot}\right)$ respectively.
\item The logarithm of Star formation Rate (SFR) and specific SFR, averaged over 100 Myr - $\log_{10}(\frac{SFR}{M_\odot \mathrm{yr}^{-1}})$ and $\log_{10}\left(\frac{\mathrm{sSFR}}{\mathrm{yr}^{-1}}\right)$.
\item The logarithm of the fraction of bolometric luminosity that is due to AGN emission using the \citet{Nenkova_A,Nenkova_B} CLUMPY models \citep{Leja_2018ApJ...854...62L} - $\log_{10}(f_{\mathrm{AGN}})$.
\item The optical depth (at 5500\,\AA{}) of a dust clump in the AGN torus (\citealt{Leja_2018ApJ...854...62L} and templates \citet{Nenkova_A,Nenkova_B}) - $\log_{10}(\tau_{\mathrm{AGN}})$.
\item The optical depth of the diffuse dust in the galaxy - $\tau_2$.
\item The optical depth of the birth cloud dust attenuation as a fraction of $\tau_2$, (only affects stars younger than 10 Myr \citep{charlot_and_fall}) - $\frac{\tau_1}{\tau_2}$.
\item The logarithm of the gas phase metallicity -  $\log_{10}\left(\frac{Z_{\mathrm{gas}}}{Z_\odot}\right)$.
\item The power law index of a modified \citet{Calzetti} dust attenuation law -  $\delta$. The power law modification is by \citet{NOll_2009}. The UV bump strength is tied to the slope following \citet{kriek_2013ApJ...775L..16K}, $\delta$;
\item The minimum starlight intensity that the dust grains are exposed to and fraction of dust mass exposed to starlight of minimum intensity - $U_{\mathrm{min}}$ and $\log_{10}(\gamma_e)$ respectively.
\item The fraction of dust mass in polycyclic aromatic hydrocarbons (PAHs) - $Q_{\mathrm{PAH}}$. (This parameter, along with the $U_{min}$ and $\log_{10}(\gamma_e)$ parameters, relate to the \citet{Draine} silicate-PAH dust emission model).
\end{enumerate}

\subsection{\texttt{Prospector-$\alpha$} SED Fit}\label{appendix:Prospector SED}
\begin{figure}[ht!]
\includegraphics[width=\textwidth]{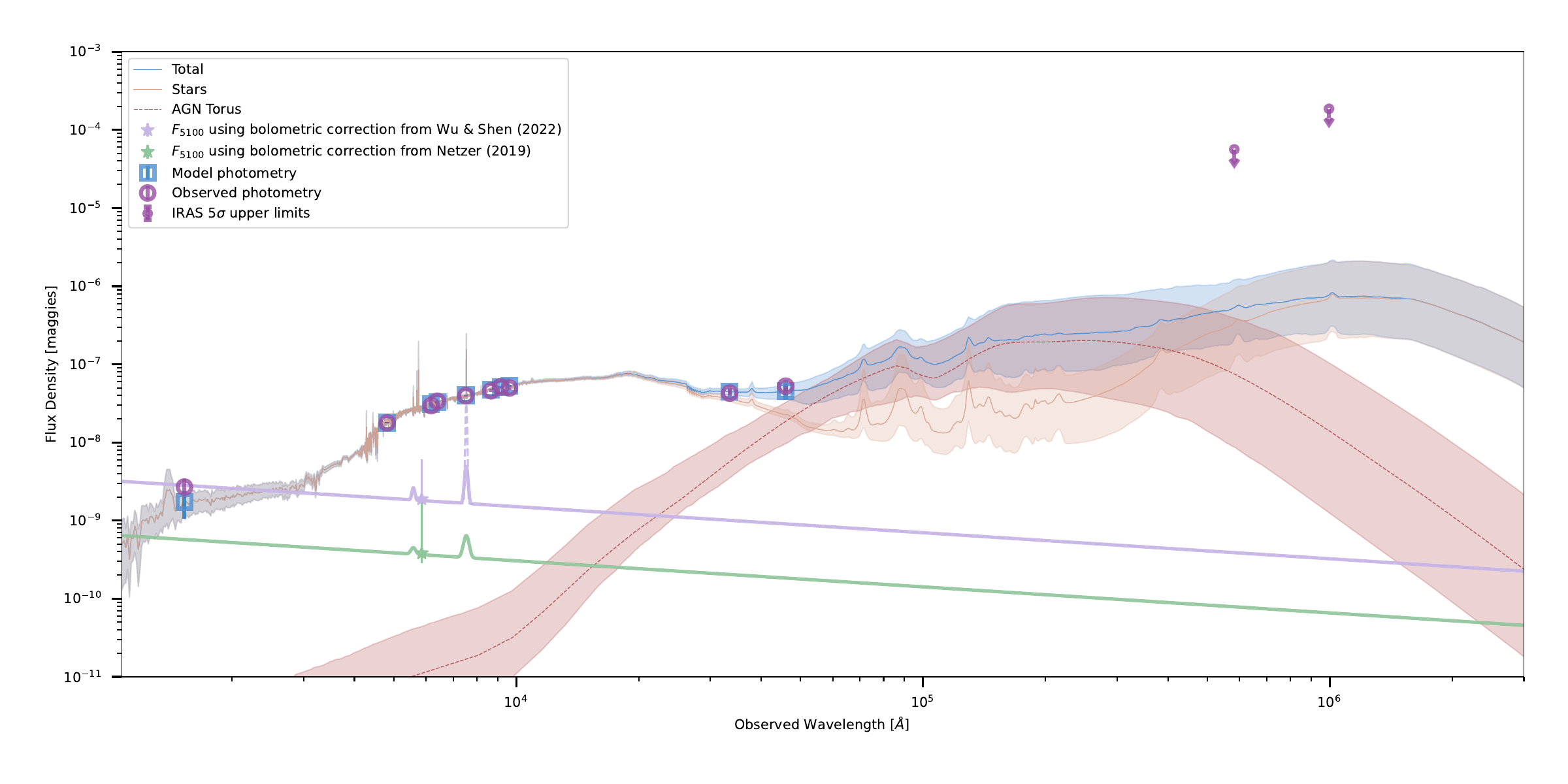}
\caption{SED plot showing the median, 16th and 84th percentile of the \texttt{Prospector-$\alpha$} fits for the total galaxy emission in blue, stellar emission in orange and the AGN torus emission in red (IRAS $5\sigma$ non-detection upper limits are shown for comparison but were not used in the fit). The $F_{5100}$ values and uncertainties derived using bolometric corrections of \citet[][green star]{Netzer_2019} and \citet{Wu_2022}, based on \citet[][purple star]{Richards_2006}, are shown together with the corresponding accretion-disk continua, assuming a power law with $\alpha=-7/3$ and plotted as solid lines of the same color. Using these $F_{5100}$ values, the corresponding broad H$\alpha$ and H$\beta$ luminosities and widths are calculated following \citet{2005ApJBHHalpha} and plotted in the same colors as their associated continua. The broad H$\alpha$ luminosity and width inferred directly from the [\ion{O}{3}] emission using the relations of \citet{Stern_Laor_2012} are shown as a dashed purple line over the \citet{Wu_2022}-based continuum.}
\label{fig:Prospector}
\end{figure}

We model the host-galaxy SED using \texttt{Prospector}, adopting the \texttt{Prospector-$\alpha$} framework \citep{Leja_2017,Johnson_2021}, which assumes a nonparametric star-formation history and includes dust attenuation, nebular emission, infrared dust re-emission, and an AGN torus component \citep{Leja_2018ApJ...854...62L}. Parameter inference was performed using the \texttt{DYNESTY} nested sampling algorithm \citep{Speagle_2020}. The model was constrained using Pan-STARRS photometry \citep{PanSTARRS2004SPIE.5489...11K} obtained from VizieR \citep{vizier}, as well as GALEX \citep{GALEX2005ApJ...619L...1M} and DES \citep{DES2016} photometry retrieved via \texttt{Blast} \citep{jones2024blastwebapplicationcharacterizing}, and NEOWISE \citep{Mainzer_2011_NEOWISE} photometry obtained through  NASA/IPAC Infrared Science
Archive \citep{NASA_IPAC2018cwla.conf...25T}. The inferred host-galaxy properties are consistent with those derived by \texttt{Blast} (Section \ref{section:blast}). 
Figure \ref{fig:Prospector} presents the photometric measurements along with the best-fit SED, and its constituent stellar emission and torus components. We also present our calculated $5\sigma$ IRAS \citep{IRAS_1984} 60 $\mu$m and 100 $\mu$m non-detection upper limits (obtained via the Scan Processing and Integration tool), which were not used in the fit, but are consistent with it. 
We plot three estimates of the expected broad H$\alpha$ emission. First, the expected broad H$\alpha$ using our [\ion{O}{3}] emission and relation from \citet{Stern_Laor_2012}.
We then derive the expected AGN continuum emission at 5100\,\AA{}, $F_{5100}$, from our estimated $L_{bol}$ using the  bolometric corrections of \citet{Netzer_2019} and \citet{Wu_2022}, based on \citet{Richards_2006}. These two $F_{5100}$ estimates, together with our average SMBH mass estimate of $\log_{10}(M_{BH})\sim7.1$, are used with the relations of \citet{2005ApJBHHalpha} to estimate the corresponding broad H$\beta$ and H$\alpha$ emission from the BLR Figure \ref{fig:Prospector} shows that both the accretion disk and expected BLR are outshined by the galaxy stellar emission in the optical bands in all cases except the \citet{Stern_Laor_2012} method.

In order to estimate the MIR luminosity of the torus, we isolate the AGN component in the \texttt{Prospector-$\alpha$} model by subtracting an SED constructed from the best fit model parameters but with the AGN contribution ($f_{AGN}$) set to zero, from the best full best-fit SED, and integrating over 1--100$\,\mu$m \citep{Stalevski_2016}. We find $L_{torus}=1.46^{+2.60}_{-0.89}\times10^{43}\, {\rm erg\,s^{-1}}$. Using the ratio $L_{torus}/L_{AGN}$ as a proxy for the torus covering factor, and adopting a torus geometry anchored to the sublimation radius \citep[Equation 5 of ][T-sampled uniformly between sublimation temperatures of Silicate and Graphite and setting $Y\equiv R_{\rm out}/R_{\rm in}$ between $10$ and $30$]{Barvianis_1987}, we infer a torus scale height of $0.59^{+0.78}_{-0.33}\ \rm pc$. 
This value is consistent with typical AGN torus sizes \citep[$\sim$0.1--10 pc;][]{Netzer2015ARA&A..53..365N,Torus_size2008ApJ...681..141R} and is roughly several orders of magnitude larger than the characteristic blackbody radii of TDEs, and also of the slightly larger radius of AT\,2022csn.

\begin{figure}[ht!]
\includegraphics[width=\textwidth]{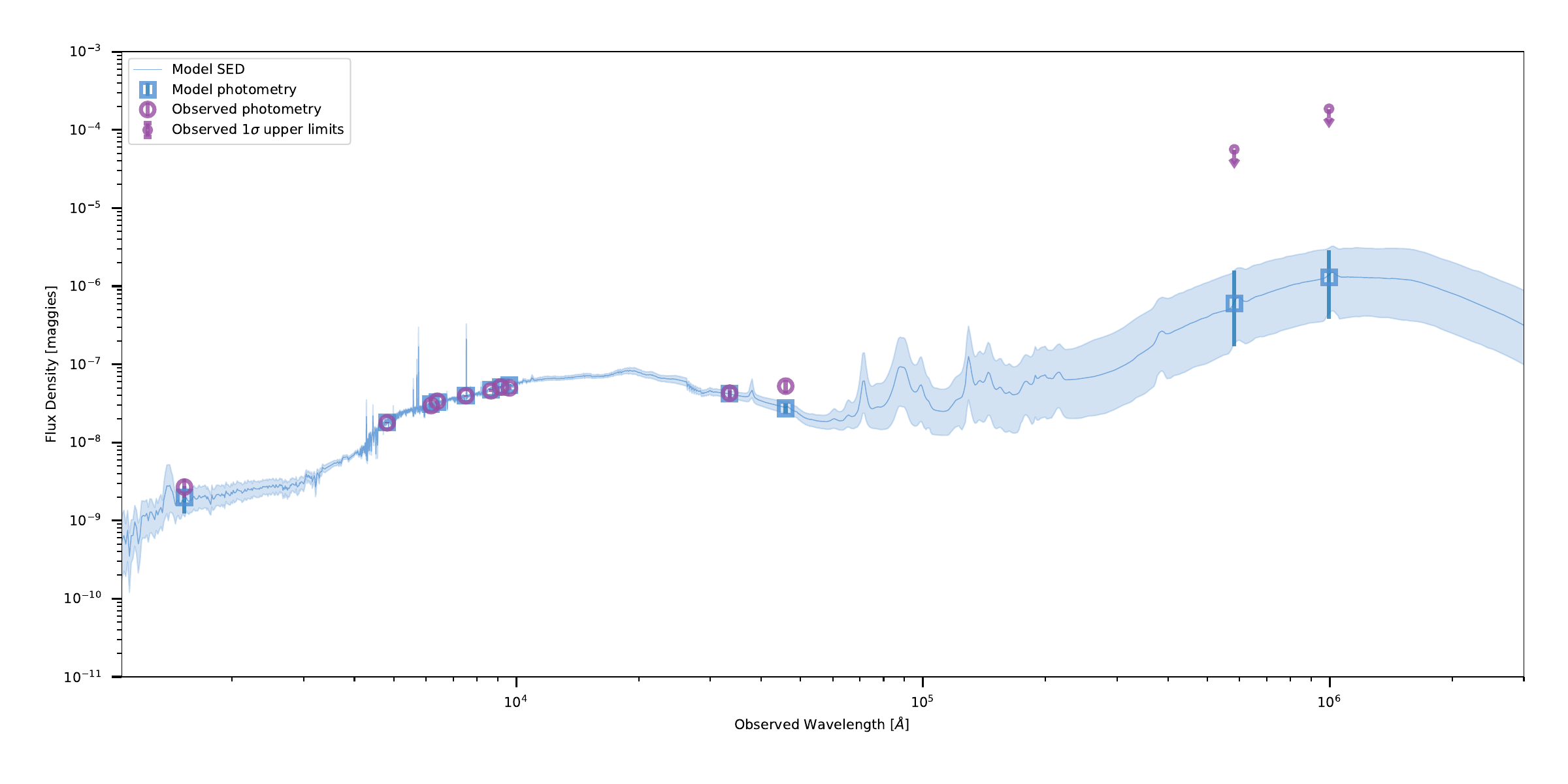}
\caption{SED plot showing the median, 16th and 84th percentile of the \texttt{Prospector-$\alpha$} fits for the total galaxy emission without an AGN component. The IRAS 5$\sigma$ upper limits were included in the fit. Stellar emission alone can not reproduce the observed $W1$ and $W2$ photometry.}
\label{fig:prospector_no_agn}
\end{figure}

To test whether the observed host-galaxy photometry, particularly in the $W1$ and $W2$ bands, could be explained without any torus contribution, we re-ran the \texttt{Prospector-$\alpha$} model using the same photometric data as before, but this time including the IRAS upper limits in the SED fitting, and with the AGN component disabled. 
As seen in Figure \ref{fig:prospector_no_agn}, the model without an AGN component cannot reproduce the WISE color as well as the model with an AGN component (Fig. \ref{fig:Prospector}).We conclude that a torus component is required in order to explain the observed $W1$ and $W2$ photometry.

\subsection{X-ray Upper limits}\label{appendix:X-ray upper limits}
Pre-flare X-ray non-detections from eROSITA \citep{erosita2021A&A...647A...1P} yield a $3\sigma$ upper limit of $L_{0.2-5.0\,{\rm keV}}\lesssim1.02\times10^{43}\ {\rm erg\,s^{-1}}$. Assuming a power-law spectrum with photon index $\Gamma=2$, we convert the upper limit to $L_{2-10\,{\rm keV}}\lesssim5.1\times10^{42}\ {\rm erg\,s^{-1}}$. Using the \citet{Ueda_2015} relation between [\ion{O}{3}] and hard X-ray luminosities, our measured $L_{[O_{III}]}=7.43\pm 1.24 \times10^{40}\ {\rm erg\,s^{-1}}$ implies $L_{2-10\,{\rm keV}}\sim1.3\times10^{43}\ {\rm erg\,s^{-1}}$, about 0.4 dex above our X-ray upper limit. This offset is smaller than the intrinsic scatter of the relation ($\sim$ 0.53 dex after inversion), and is therefore consistent with the local hard X-ray selected AGN sample of \citet{Ueda_2015}.

\section{Fits to the Host Galaxy Spectroscopy}\label{appendix:host_galaxy_spectral_fits}

\begin{figure*}[ht!]
\includegraphics[width=\textwidth]{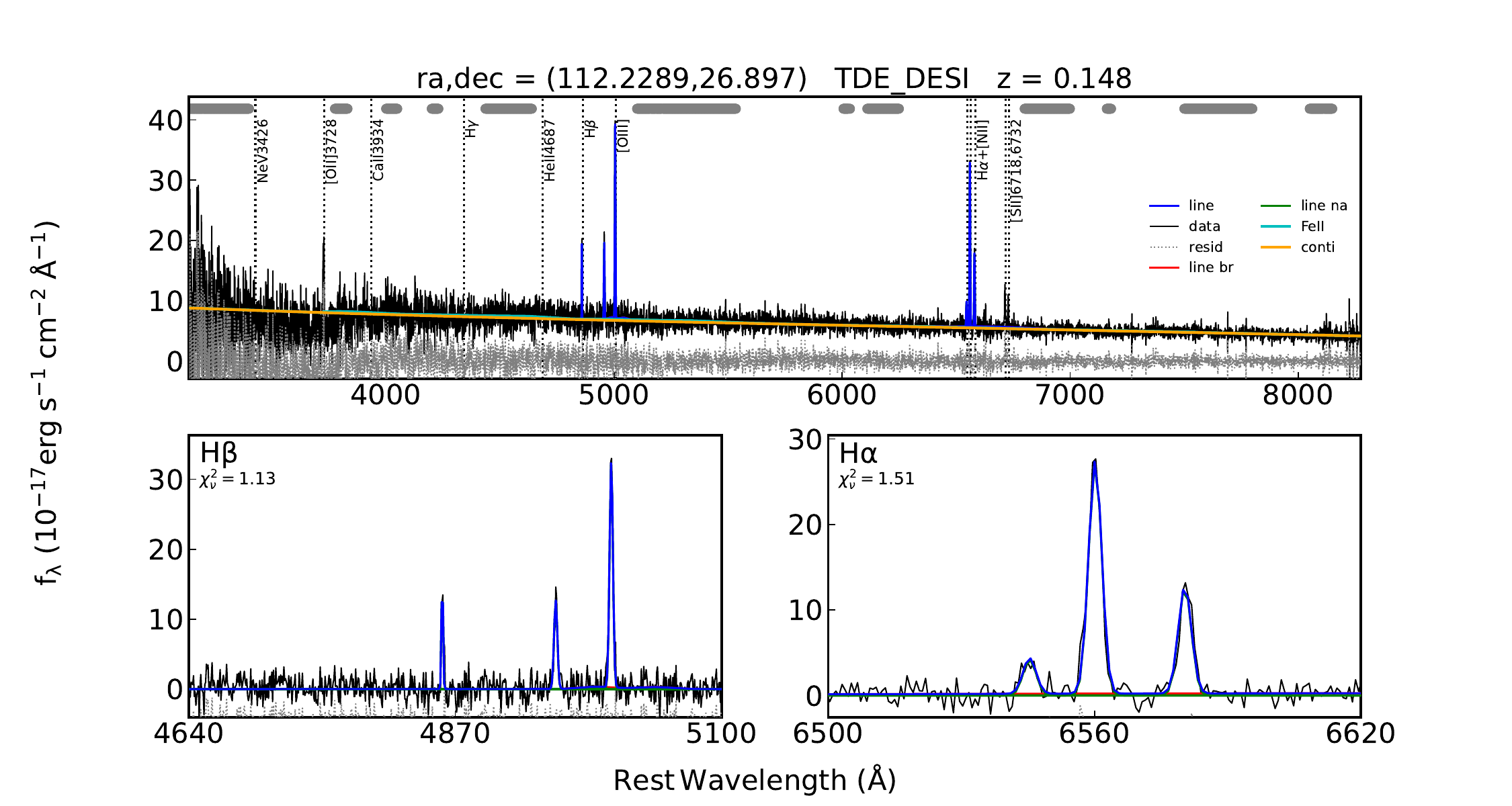}
\caption{PyQSOFit fit to the DESI host-galaxy Spectrum. The total emission-line model is shown in blue, with the narrow and broad line components in green and red, respectively. The \ion{Fe}{2} emission template is shown in cyan, the continuum in orange, and the residuals are depicted by the black dotted line.}
\label{fig:PyQSOFit_DESI}
\end{figure*}

Figure \ref{fig:PyQSOFit_DESI} presents the \texttt{PyQSOFit} continuum and emission-line model fit to the DESI spectrum of the host galaxy obtained $\sim$28 days prior the discovery of the TDE. The fit reproduces the H$\alpha$, H$\beta$, [\ion{O}{3}], and [\ion{N}{2}] emission lines well, but provides a poorer match to the [\ion{S}{2}] $\lambda\lambda$6716,6731 doublet.

\section{BPT Diagram with Sulfur doublet subtraction}\label{appendix:BPT_2}
\begin{figure}[ht!]
\centering
\includegraphics[width=0.7\textwidth]{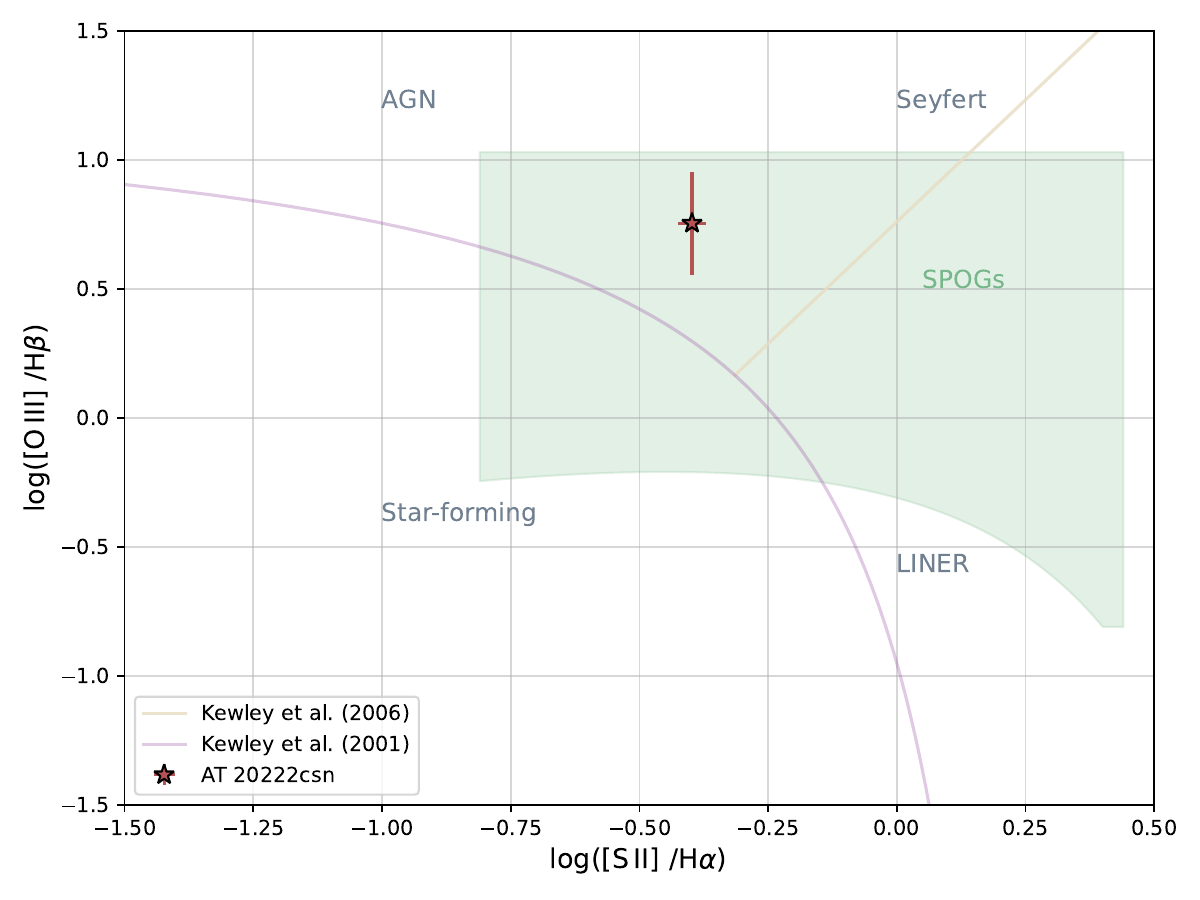}
\caption{Same as Figure \ref{fig:BPT} but using the [\ion{O}{3}]/H$\beta$ and [\ion{S}{2}] $\lambda\lambda$6716,6731/H$\alpha$ line-flux ratios, and the division between Seyferts and LINERs of \citet{Kewley_2006}.}
\label{fig:BPT_SII}
\end{figure}

Figure \ref{fig:BPT_SII} shows the BPT diagram for the host of AT\,2022csn, using the [\ion{O}{3}]/H$\beta$ and [\ion{S}{2}] $\lambda\lambda$6716,6731/H$\alpha$ line-flux ratios measured from the host-galaxy DESI spectrum. Since the [\ion{S}{2}] doublet is not well captured by automated fitting with \texttt{PyQSOFit}, and there are no broad emission components present in the host-galaxy lines (Figure \ref{fig:Stacked_spectra_norm}), we measure the narrow line fluxes by fitting the local continuum and integrating the emission above it. The host lies outside the star-forming locus, in the AGN-Seyfert region, and is also consistent with the SPOGs population \citep{Altalo_2016ApJS..224...38A}.

\begin{figure}[ht]
\centering    \includegraphics[width=0.7\textwidth]{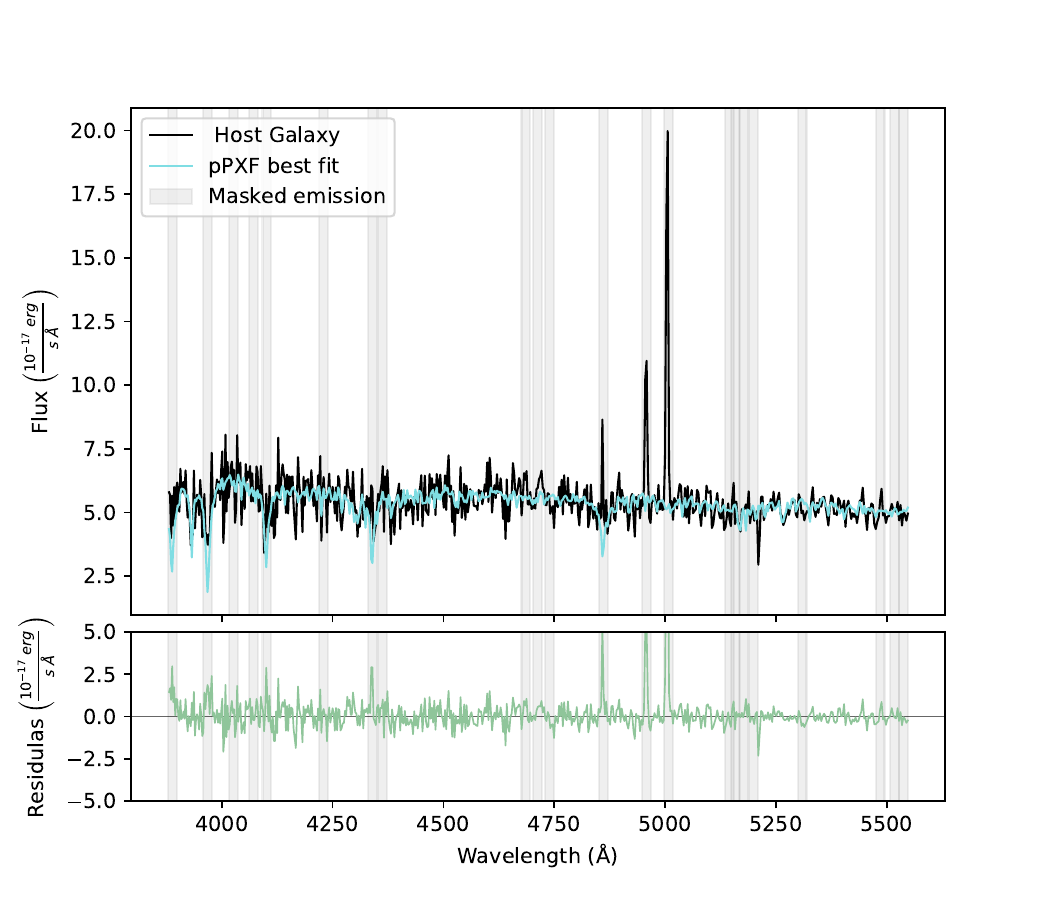}
\caption{\texttt{pPXF} best-fit galaxy model to the DESI host-galaxy spectrum. The observed spectrum is marked in black (binned here for clarity, though the fit was performed on the full data), the best fit model in cyan and the residuals of the fit in green.
Areas marked in grey denote wavelength regions of emission lines masked from the fit following \cite{Koss_2022}.}
\label{fig:ppxf}
\end{figure}

Figure \ref{fig:ppxf} shows the \texttt{pPXF} best-fit stellar model to the DESI host-galaxy spectrum.

\clearpage

\bibliography{sample701}{}
\bibliographystyle{aasjournalv7}

\end{document}